\documentclass[acmlarge]{acmart}

\makeatletter
\newcommand{\myconfshort}{\acmConference@shortname}
\newcommand{\myconffull}{\acmConference@name}
\newcommand{\myconfdate}{\acmConference@date}
\newcommand{\myconfloc}{\acmConference@venue}
\AtBeginDocument{
  \fancypagestyle{firstpagestyle}{
    \fancyhead{}%
    \fancyfoot[C]{}%
  }
  \fancyhf{}
  \fancyhead[LO]{\@headfootfont\shorttitle}%
  \fancyhead[RE]{\@headfootfont\@shortauthors}%
  \fancyhead[LE]{\@headfootfont\footnotesize \myconfshort, \myconfdate, \myconfloc}%
  \fancyhead[RO]{\@headfootfont\footnotesize \myconfshort, \myconfdate, \myconfloc}%
  \fancyfoot[C]{}%
}
\makeatother
\acmBooktitle{\conffull\@ (\confshort), \confdate, \confloc}
\AtBeginDocument{%
  }

\setcopyright{cc}
\setcctype{by} 
\copyrightyear{2026}
\acmYear{2026}
\acmDOI{10.1145/3805689.3812388}
\acmConference[FAccT '26]{The 2026 ACM Conference on Fairness, Accountability, and Transparency}{June 25--28, 2026}{Montreal, QC, Canada}
\acmBooktitle{The 2026 ACM Conference on Fairness, Accountability, and Transparency (FAccT '26), June 25--28, 2026, Montreal, QC, Canada}
\acmISBN{979-8-4007-2596-8/2026/06}

\usepackage{tcolorbox}
\usepackage{enumitem}

\definecolor{boxborder}{RGB}{0, 0, 0}
\definecolor{boxbg}{RGB}{248, 248, 248}
\definecolor{headerbg}{RGB}{220, 220, 220}

\newtcolorbox{plausibilitybox}[2][]{
  colback=boxbg,
  colframe=boxborder,
  fonttitle=\bfseries,
  coltitle=white,
  title={#2},
  sharp corners,
  boxrule=0.8pt,
  #1
}

\newcommand{\chksection}[1]{%
  \par\vspace{4pt}\noindent\colorbox{headerbg}{\parbox{\dimexpr\linewidth-2\fboxsep}{\textbf{\small #1}}}\par\vspace{2pt}%
}

\newcommand{\cbox}{\makebox[0pt][l]{$\square$}\raisebox{.15ex}{\hspace{0.1em}$\checkmark$}}

\begin{document}

\title{Mechanism Plausibility in Generative Agent-Based Modeling}

\author{Patrick Zhao}
\email{patrick_zhao@sfu.ca}
\orcid{0009-0000-5494-868X}
\affiliation{%
  \institution{Simon Fraser University}
  \city{Burnaby}
  \state{BC}
  \country{Canada}
}

\author{David Huu Pham}
\email{dhpham@sfu.ca}
\orcid{0009-0007-6426-7467}
\affiliation{%
  \institution{Simon Fraser University}
  \city{Burnaby}
  \state{BC}
  \country{Canada}
}

\author{Nicholas Vincent}
\email{nvincent@sfu.ca}
\orcid{0000-0002-8493-7161}
\affiliation{%
  \institution{Simon Fraser University}
  \city{Burnaby}
  \state{BC}
  \country{Canada}
}

\renewcommand{\shortauthors}{Zhao et al.}

\begin{abstract}

Large language models (LLMs) can generate high-level diverse phenomena without explicitly programmed rules. This capability has led to their adoption within different agent-based models (ABMs) and social simulations. Recent research aim to test whether they are capable of generating different phenomena of interest, for example, human behavior on social media platforms or performance in game-theoretic scenarios.

However, capability, prediction, and explanation are different -- drawing from the philosophy of science and mechanisms literature, \textit{explanation} requires showing, to some degree, how a phenomenon is produced by related organized entities and activities. For modelers, describing the characteristics of an experiment or whether a simulation provides progress in capability (or explanation), can be difficult without being grounded in potentially distant research areas.

We integrate recent work on LLM-ABMs with contemporary philosophy of science literature and make two main contributions. First, we gather insights from modeling and mechanisms literature and use them to operationalize a definition of `plausibility' in a four-level scale. Our scale separates the evaluation of a model's generative sufficiency (ability to reproduce a phenomenon) from its mechanistic plausibility (how the phenomenon could be produced), and clarifies the distinct roles of different models, such as predictive and explanatory ones. We introduce this as the Mechanism Plausibility Scale. Second, we discuss the early wave of LLM-ABM research and find that papers often conflate evidence of Agent-level functionality with claims about emergent ABM-level phenomenon, relying on `believability' metrics that focus on generative sufficiency. Our discussion section speaks on how these findings echo long-standing problems in classical ABM, historical harms caused by these issues, and broader ethical and epistemic concerns about using LLMs in modeling. Using the findings from our review, we offer the scale as a practical heuristic in the form of a checklist which can clarify how simulations at different levels of plausibility may be useful. We hope the activity of filling out the scale will help new modelers ground the epistemic contribution of their simulations. 

\end{abstract}




\maketitle


\section{Introduction}

Developments in natural language processing have spurred interest in using large language models (LLMs) in social simulations \cite{hortonLargeLanguageModels2023, parkGenerativeAgentsInteractive2023, anthisLLMSocialSimulations2025}, for example, extending the action space of agents in agent-based models (ABMs). It seems increasingly possible that product decisions, policymaking, and even research itself may be influenced by the outcome of such simulations \cite{anthisLLMSocialSimulations2025, liWhatMakesLLM2025}. Today, a modeler \cite{weisbergWhoModeler2007} creating simulations with LLMs is capable of reproducing higher-level phenomena without explicitly programming the mechanisms that produce them. In a canonical agent-based model (ABM), the mechanisms underlying a phenomenon are operationalized and programmed by the modeler (e.g., for an economic agent, ``if the price of resource $A$ is $X$ and my internal requirement $B$ is at a threshold $Y$, then buy some amount of resource $A$''). These rules are typically based on some combination of assumptions, scientific theory, and empirical data. For instance, a programmer might read several social science papers to determine a particular distribution from which their agents will draw values that represent their preferences and attributes. Using LLMs in simulations offers the tantalizing promise that weights and biases obtained by training on social data may contain relevant distributional information about human behavior, allowing for richer representations of human subjects \cite{vanheeLargeLanguageModels2025, anthisLLMSocialSimulations2025, parkGenerativeAgentsInteractive2023}. On the other hand, critiques have also formed around models failing to capture the complete experiences of the human subjects they substitute \cite{agnew_illusion_2024}, which leaves us with questions about if this is tied to the nature of LLMs, and if so, the question of if LLMs should be used at all.

When using LLMs in modeling social phenomena, we are left with a few puzzles: For a given simulation, did the results emerge from some correctly retrieved social knowledge encoded in the LLM's weights? Do our agents model the human behavior we are interested in? This could be the case, given that LLMs are trained on data describing real, human decision-making. Without improvements in the field of machine learning (ML) interpretability and data attribution, it could be the case that simulations incorporating LLMs are drawing on information that is irrelevant to the modeler's intent (sometimes referred to in the machine learning space as `faithfulness' \cite{yonaCanLargeLanguage2024, paulMakingReasoningMatter2024}). In other words, we might produce a simulation that looks like it is \textit{explaining} a social science phenomenon, but is just generating it through some other means, regardless of the `how'. One can imagine that there are many ways in which a phenomenon can occur, and we are only interested in a particular one.

A recent review by Larooij et al. from April 2025 \cite{larooijLargeLanguageModels2025} surveyed and found that a number of studies involving ABMs with LLMs (LLM-ABM) fail to acknowledge established work in the traditional simulation literature, or even have proper operational validity. One particular summary is that recent evaluations rest on some variant of believability, where human annotators are tasked with labeling whether or not they think the outputs of agent dialogues are produced by a human. On top of this, much work focuses on whether or not a simulation or its LLM agents are capable of producing a specific phenomenon.

These problems leave us with further questions: Is it necessary to completely understand the inner workings of LLMs to produce useful simulations? What does `useful' mean anyways, in the context of simulations? To facilitate the discussion we connect work from the philosophy of science about what can be learned from idealized computer models, such as ABMs. 

Let us consider a target phenomenon $T$ a modeler is attempting to produce using a simulation $S$. In traditional agent-based modeling it is mostly accepted that by generating $T$ using $S$, they realize a possible candidate for how $T$ is created, sometimes called generative sufficiency \cite{epsteinGenerativeSocialScience2006}. By `growing' $T$ through their simulation, the modeler has created an input-output mapping and demonstrated a sufficient, but not necessary condition for how $T$ might arise \cite{epsteinGenerativeSocialScience2006}. 



Now suppose the modeler wants to explain, to some level, how $T$ arises in actuality--they need to describe the relationship between the simulation's mechanisms and the ``real'' mechanisms that produce the target. \textit{Mechanisms} are the theoretical organization of entities and activities behind a phenomenon \cite{machamerThinkingMechanisms2000, craverMechanismsScience2024}. In the mechanisms and neuroscience literature a simulation that produces $T$ without a connection to the ``how-actually'' is called a \textit{phenomenal} model \cite{kayPrinciplesModelsNeural2018, maukPotentialEffectivenessSimulations2000}. Connecting this to ABMs, a modeler might use a simulation to deduce or intuit parts of the mechanisms behind $T$; however, since a simulation of $T$ is only a single possible candidate, one could say that generative success is not sufficient to show the mechanisms in $S$ correspond to the mechanisms responsible for $T$. A modeler could generate $T$ in many possible ways, perhaps completely unrelated to any hypothesis about its real causes. If a modeler is interested in creating a simulation that helps in explaining the target phenomenon, it needs to convey some level of information about the underlying mechanism and propose how it is mapped to the simulation; that is, they need to show the mechanisms in $S$ are \textit{plausible} mechanisms for $T$. 

In this paper we explain how simulations can vary in their level of plausibility and introduce a set of criteria for categorizing simulations along our axis of interest. This is not to say that the value of a simulation is dictated purely by plausibility or the mechanistic understanding of a model; it is of general agreement that idealizations and abstractions are common, if not, necessary in building accepted models, or science may never move forward \cite{elginTrueEnough2004, aydinonatPuzzleModelbasedExplanation2024a, parkerModelEvaluationAdequacyforPurpose2020}. Simulations can vary in their level and type of claim, whether they claim to be predictive, illustrative, exploratory, explanatory, etc. However, if a modeler wants to use their simulation to make any level of claim about how $T$ might arise in actuality (explanatory), they must move beyond a purely phenomenal account. 
We present a checklist version of the scale in Section \ref{sec:ApplyingScale}, motivated by dataset and model information checklists in past work \cite{gebruDatasheetsDatasets2021, mitchellModelCardsModel2019a, winikoffScoresheetExplainableAI2025}. We believe the scale will guide researchers in developing their own models, especially those integrating LLMs.

In Section \ref{sec:Definitions} we operationalize and elaborate on terms used across the paper such as \textit{mechanism}, \textit{phenomenal}, \textit{plausibility}, and \textit{explanation}. We aim to show how these concepts can be directly related to existing simulation work in various fields of computing, especially the use of LLM simulation across human computer interaction and computational social science. In Section \ref{sec:PlausibilityLevels} we introduce a ``mechanism plausibility scale'' that aims to capture core ideas from the diverse literatures discussed in the preceding section and provide a practical approach for classifying simulations and their contributions. In Section \ref{sec:ApplyingScale} we present the more pragmatic, checklist version of the scale and discuss the reviews involved in its development. Finally, in Section \ref{sec:LLMInSocialSimulation} we further discuss contemporary problems of LLM-enabled simulation and how it relates to their placement on these scales. 


\section{Motivation and Related Work}\label{sec:Definitions}

\subsection{Operationalization}
\label{subsec:Operationalization}


In the philosophy of science, \textit{phenomena} are defined to be stable patterns, regularities, or events that can be reliably inferred from data, and are the targets of explanation for scientific communities \cite{bogenSavingPhenomena1988}.
The patterns that qualify as `phenomena' are scoped to the particular domain of inquiry \cite{massimiPerspectivalOntologySituated2022}, and may vary depending on the modeler's methodological choices or research question \cite{mcallisterPhenomenaPatternsData1997}.

Consider a subject who displays the phenomena of eye contact avoidance and shaking limbs. The phenomena of `eye contact avoidance' is something that is inferred by patterns in the data pertaining to the subject's average length of eye contact and their direction of gaze. Although behavioral, psychological, or social phenomena may be inferred from aggregated, third-person observational data, the mental experience of, and the cause of these phenomena may only be accessible to the subject experiencing them \cite{titchenerTextbookPsychology1910, jacksonEpiphenomenalQualia1982}, where outside observers can only agree upon a subject's apparent reactions to their own internal experience \cite{stevensOperationalDefinitionPsychological1935}. Third-person observers may posit that the phenomena displayed by the subject are indicative of the \textit{hypothetical construct} of anxiety \cite{maccorquodaleDistinctionHypotheticalConstructs1948, cronbachConstructValidityPsychological1955}. 


Originating from psychometric evaluation, hypothetical constructs are a theoretical attribute postulated to explain observed behavioral patterns, but are not directly observable themselves \cite{maccorquodaleDistinctionHypotheticalConstructs1948, cronbachConstructValidityPsychological1955}.
We often work with hypothetical constructs in order to characterize and reason about mental and social phenomena \cite{vessonenConceptualEngineeringOperationalism2021, pichlerConceptsTextsBack2022}.
To allow empirical measurements of these constructs, we create operational definitions: these are an explicit, unambiguous set of operations, protocols, or rules that are treated as equivalent to these abstract constructs within the bounds of the experiment, for the sake of falsifiable detection and experimental reproducibility \cite{bridgmanLogicModernPhysics1927}.
The process of creating an operational definition for a particular concept is called the ``operationalization'' of the concept, and determining whether this operational definition measures what it is intended to measure, is called construct validity \cite{cronbachConstructValidityPsychological1955, carminesReliabilityValidityAssessment1979}. 

Operationalization involves not only translating abstract constructs into measurable patterns, but also assigning interpretations to the formal components of a model. From the modeling and cognitive representation literature \cite{weisbergSimulationSimilarityUsing2013, eganDeflatingMentalRepresentation2025a}, we recognize the distinction between a model's formal functions and the interpretation assigned by the modeler to connect the functions to the domain of interest. 
Egan distinguishes between what she calls the ``theory proper'' of a computational model and an ``intentional gloss'' that accompanies it \cite{eganDeflatingMentalRepresentation2025a}. The theory proper specifies the mathematical function(s) computed, the algorithms, the structures maintained, and their physical realization. For LLM Agents within a simulation, this would be the next-token probability distribution over a vocabulary, given an input sequence. The intentional gloss is what connects the computation to the modeler's interpretation, which could be some target persona the modeler says the agent is representing. But as Egan argues, the validity of an intentional gloss is not guaranteed by its computations and requires independent justification, typically grounded in the theorist's explanatory goals. In our scale interpretation takes the form of an Intent $I$, which we will return to in Section \ref{subsec:level2}.

Since our focus is on simulation models created by researchers, most researchers have particular phenomena that they would like the simulation to produce \cite{graebnerHowRelateModels2018}.
We refer to these phenomena of interest as $T$, the \textit{target phenomena}, and we expand on this definition in Section \ref{subsec:Phenomeonal}.
In our review we find gaps in the operationalization of target phenomena used in the evaluation of recent LLM-based social simulations, further discussed in \ref{subsec:Reflections}.

\subsection{Phenomenal Models and Generative Sufficiency}
\label{subsec:Phenomeonal}

The term \textit{phenomenal} comes from established usage in the scientific modeling field \cite{maukPotentialEffectivenessSimulations2000, kayPrinciplesModelsNeural2018}, where it is used to describe models that aim to produce the patterns describing a target phenomenon $T$, but do not contain information about the mechanism behind it. They may produce an accurate output without describing the relevant internal structure, therefore limiting their explanatory power. 

One can imagine that it is not always simple or practical to produce the target phenomenon; In the modeling field, the term \textit{generative sufficiency} describes the level to which a model is able to accurately produce $T$ \cite{graebnerHowRelateModels2018}. Due to the black-box nature of deep neural networks and other practicality reasons, much of the emphasis in the traditional machine learning field is put on the generative sufficiency of different ML models--how accurately they are able to produce a target behavior. At the time of writing, the primary ways to evaluate LLMs are to measure scores they achieve on some benchmark centered around human evaluation or fact-checking. This mentality may have spread to the LLM social simulation area, as initial projects in the space used similar evaluations to benchmark the realism of their simulation. For example, projects using `believability' as a metric for their sufficiency in producing a target behavior \cite{parkGenerativeAgentsInteractive2023, kaiyaLyfeAgentsGenerative2023, wangUserBehaviorSimulation2025, parkSocialSimulacraCreating2022, liTheoryMindMultiAgent2023, renEmergenceSocialNorms2024}. 

While generative sufficiency could be appropriate for exploratory or illustrative settings, the goals of a model may not be limited to just reproducing the target behavior; One may want to test unknown counterfactual scenarios or interventions. Problematically, these simulations are evaluated based off of their generative sufficiency and then used to test interventions as if there are plausible mechanisms \cite{huaWarPeaceWarAgent2024, gaoS3SocialnetworkSimulation2025}. Phenomenal models cannot be used to test counterfactual scenarios because they lack the relevant internal causal structure. In order to move past being purely phenomenal, the model needs to suggest how $T$ is produced: the mechanisms behind it. 

\subsection{Mechanisms}
\label{subsec:Mechanisms}

Mechanisms literature has seen a rise in discussion in the past two decades, primarily in the philosophy-of-science and neuroscience fields \cite{craverMechanismsScience2024, glennanNewMechanicalPhilosophy2017, machamerThinkingMechanisms2000}. Glennan gives a minimal definition for mechanisms in his text, \textit{The New Mechanical Philosophy:} 
\begin{quote}
    \textit{``A mechanism for a phenomenon consists of entities (or parts) whose activities and interactions are organized so as to be responsible for the phenomenon.''} \cite{glennanNewMechanicalPhilosophy2017}
\end{quote}
Pragmatically, in our discussion of agent-based models and computer simulation, this might include how the agents, environment, and update rules function to produce $T$. 
A mechanism is involved in the causal process of $T$, not just correlated, and hypothesizing about them is the first step towards explaining a phenomenon. This hypothesis can take the form of a mapping which details what parts of the simulation correspond to mechanisms behind $T$. In our scale the addition of this mapping is what distinguishes a purely phenomenal model from one that presents the plausible candidate mechanisms behind $T$. 

To explain the mechanisms behind a phenomenon is to explain how the phenomenon is produced (falsifiably). Once some level of description of the mechanisms behind $T$ are produced, the model is beyond a purely phenomenal account. Kaplan and Craver have summarized these demands into an account called the 3M requirement:
\begin{quote}
    \textit{``In successful explanatory models in cognitive and systems neuroscience (a) the variables in the model correspond to components, activities, properties, and organizational features of the target mechanism that produces, maintains or underlies the phenomenon, and (b) the (perhaps mathematical) dependencies posited among these variables in the model correspond to the (perhaps quantifiable) causal relations among the components of the target mechanism.''} \cite{kaplanExplanatoryForceDynamical2011}
\end{quote}
To be clear, an explanation does not need to constitute every detail down to the atomic level; it can use an adequate level of abstraction or idealization to fit the use case of the modeler \cite{swarupAdequacyWhatMakes2019, craverWhenMechanisticModels2006}. For example, in Figure \ref{fig:CraverDiagram}, lower-level mechanisms beyond the `Agent-level' could be further explored and abstracted, but may be stubbed at the modeler's adequate level of abstraction.

We note that mechanisms cannot be identified in isolation, and therefore the target phenomena need to be operationalized before identifying any mechanisms. Craver suggests, ``mechanistic explanations can fail because one has tried to explain a fictitious phenomenon, because one has mischaracterized the phenomenon, and because one has characterized the phenomenon to be explained only partially.'' \cite{craverExplainingBrain2009} Mechanisms are not just `static' concepts, they are functions that are defined relative to a phenomenon. Its identity, boundaries, and relevance are all defined by the specific outcome it is supposed to explain \cite{craverMechanismsScience2024, glennanMechanismsNatureCausation1996}. Therefore, as we will see later, in our Mechanism Plausibility Scale the operationalization precedes the hypothesis.

\subsection{Mechanisms vs Prediction}
The focus on the productive process of a phenomenon distinguishes mechanisms from predictivism or correlation. One can use a barometer reading to predict weather, but the changing air pressure it measures is not the mechanism that produces it. In causal inference, this is the difference between observational and interventional questions \cite{pearlCausality2009, 10.5555/3238230}. A predictive model is appropriate for questions such as ``given these initial conditions, what outcomes are likely?''  
Without plausible mechanisms, however, a model's predictive outputs are usually only appropriate to the extent that they can be validated against observed outcomes. For scenarios that have been or can be empirically tested, this validation may suffice \cite{shmueliExplainPredict2010}. But for novel interventions, mechanisms provide the basis for reasoning about whether the model's outputs are acceptable.


This distinction between explanation and prediction is well established and known to be easily conflated \cite{shmueliExplainPredict2010}. The two goals require different criteria for model evaluation, different relationships to the underlying data-generating process, etc. An explanatory model aims to test causal hypotheses about the process producing $T$; a predictive model aims to produce accurate forecasts of new observations, and may do so using variables that have no/weak causal relationship to the outcome. Conflating the two is a category error that appears in both classical statistics and, as we argue, in early LLM-ABM work.

\subsection{Plausibility}
The definition of plausibility can be vague, subjective and is often treated as a qualitative property. A general definition from the Stanford Encyclopedia of Philosophy: \textit{``To say that a hypothesis is plausible is to convey that it has epistemic support: we have some reason to believe it, even prior to testing.''} \cite{bartha_analogy_2024}

In this paper, we operationalize plausibility in ABMs based off of its standing on our scale (presented in Section \ref{sec:PlausibilityLevels}), which encapsulates factors such as how a simulation's components are operationalized, the type of evidence used to justify its parameters, the model's relationship to hypotheses the modeler is presenting, etc. In particular, we are interested in if a model is a faithful representation of the modeler's intent. As mentioned previously, scholars recently publishing in the LLM-ABM space have used believability/plausibility metrics like human annotation to support their simulations. We find that the task of identifying what these evaluations actually provide support for is elusive for even seasoned and capable researchers when it is applied to LLM simulation, giving us the primary motivation for developing our scale. 

\begin{figure*}[t]
    \centering
    \includegraphics[width=0.7\linewidth]{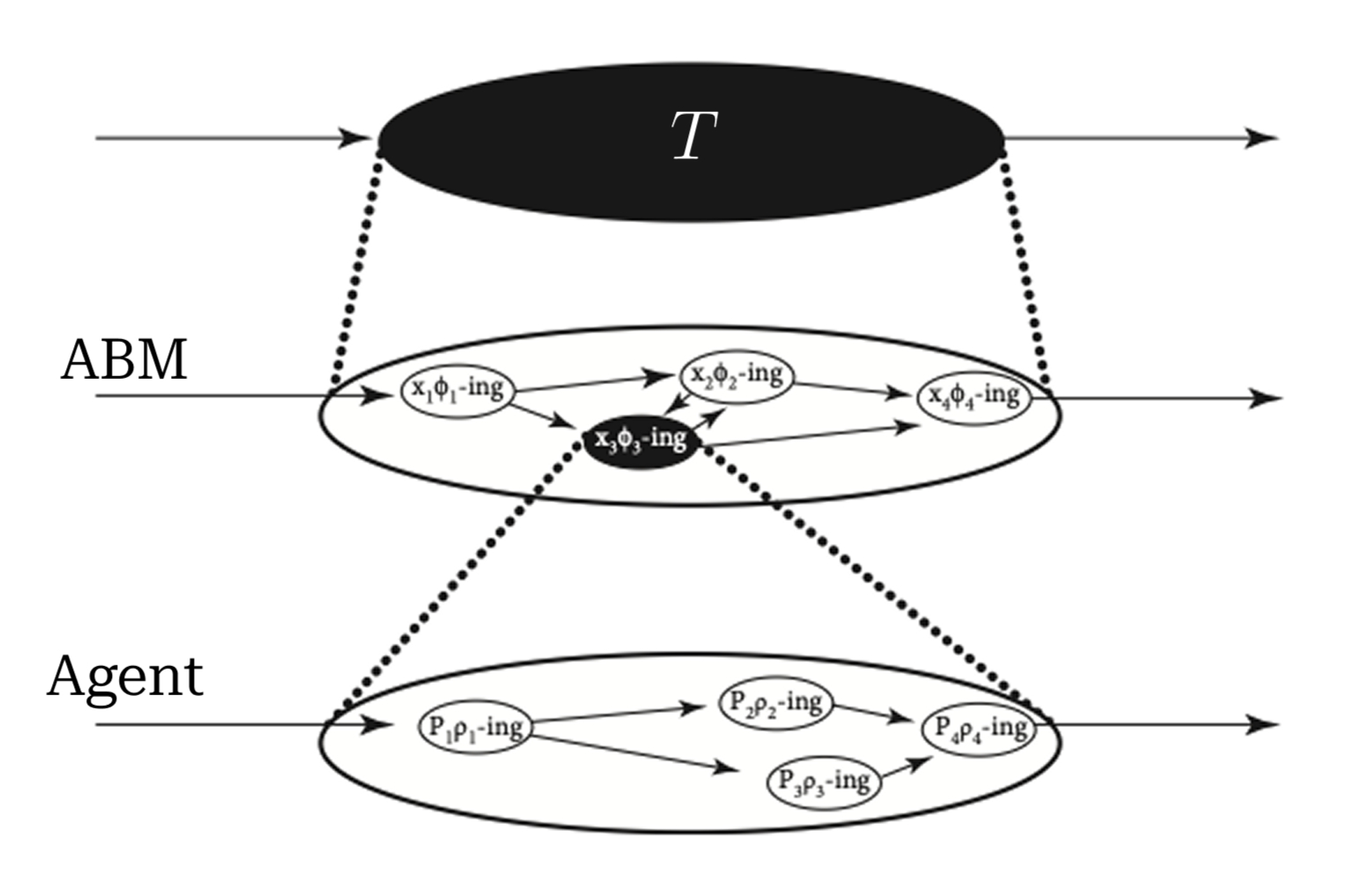}
    \caption{
        An adapted Craver diagram \cite{craverExplainingBrain2009} showing a simulation producing $T$ with higher and lower-level mechanisms. In the ABM, the agents/entities $\{x_1,\ldots,x_m\}$ (circles) and activities $\{\phi_1,\ldots,\phi_n\}$ (arrows) work to produce $T$. The agents in the ABM are further and reciprocally constituted by lower-level mechanisms, which are generally abstracted away for the purposes of tractability, but are also why simulations can never be fully validated (see Level $\Omega$ in Section \ref{subsec:PlausibilityClarifications}).
    }
    \label{fig:CraverDiagram}
\end{figure*}

\section{A Mechanism Plausibility Scale}\label{sec:PlausibilityLevels}

Now that we have distinguished explanatory models from predictive and illustrative ones, we introduce an axis for models as plausible explanations.

\begin{quote}
    Craver: \textit{``For those interested in building plausible simulations, it will not suffice for simulation $S$ simply to reproduce the input–output mapping of target phenomenon $T$. The model is further constrained by what is known about the internal machinery by which the inputs are transformed into outputs. It is possible, for example, to simulate human skills at multiplication with two sticks marked with logarithmic scales; but that is not how most humans multiply.''} \cite{craverWhenMechanisticModels2006}
\end{quote}

If we want a simulation to be a \textit{plausible} representation for how $T$ is created, it is not sufficient to just reproduce $T$ -- the simulation mechanisms must be adequately close to being a proxy for how $T$ may actually be generated \cite{kaplanExplanatoryForceDynamical2011, weisbergSimulationSimilarityUsing2013}. 

We formalize a \textbf{model} as a four-tuple $M=(S,T,I,E)$: \textbf{Simulation} ($S$), \textbf{Target} phenomenon ($T$), modeler \textbf{Intent} ($I$), and \textbf{Evidence} ($E$) (these terms will be further defined below). We use this four-tuple to create a corresponding four-level \textbf{Mechanism Plausibility Scale}. Models climb our scale as more components of $M$ become falsifiable and relevant in explaining its mechanisms, as well as being overall more faithful in representing the modeler's intentions. To make the scale concrete, we will revisit at each level a high-level running example: a (hypothetical) LLM-based simulation of opinion dynamics on a social media platform.

\subsection{Level 0}
\label{subsec:level0}
A Level 0 model is a `toy' simulation or sandbox with no specified modeling goal. It consists of a Simulation $S$, which is set of procedures, code, and update rules that generate outputs but lacks a clearly defined phenomenon to explain. Since mechanisms are defined relative to a phenomenon (sometimes referred to as \textit{Glennan's Law}) \cite{glennanMechanismsNatureCausation1996, craverMechanismsScience2024}, a model without a target cannot have mechanisms. We place models that lack explicit operationalization of a target $T$ unintentionally in level 0 as well. 


\subsubsection{Example} A research team builds a sandbox where LLM agents are placed on a simulated social network and allowed to post, reply, and share content freely. The purpose is to demo and explore a new simulation technique. The researcher documents the system and observes what happens, but has no specified phenomenon to reproduce or explain. 

\subsection{Level 1}
\label{subsec:level1}
To reach Level 1, a model must add an \textbf{operationalized} target $T$. 

Models that do not convey anything about the underlying mechanism are said to be \textit{phenomenal}, their purpose being pattern reproduction rather than creating hypotheses with their simulation \cite{craverWhenMechanisticModels2006, craverMechanismsScience2024}. Models at level 1 are \textit{phenomenal}; They have an operationally defined $T$ and are considered generatively sufficient if its simulation $S$ can produce the operationalized patterns of $T$. However, it makes no claims about explanation. $S$ exists to produce $T$ in any way. To put it another way, they are `hard-coded' simulations that produce a set of data points which match a pattern operationalized as $T$. 

Recently, Level 1 models using LLM agents have been used to explore the capabilities of different language models in cooperation, games, and other environments. The goal is not to accurately model the scenario, but to benchmark how different LLMs perform in those abstract environments. For example, there exists a multitude of work on placing LLMs in game-theoretic scenarios to see how they act or perform \cite{akataPlayingRepeatedGames2025, costarelliGameBenchEvaluatingStrategic2024, guoGPTGameTheory2023, kempinskiGameThoughtsIterative2025, li_spontaneous_2025}. 

The research questions of these simulations often follow the lines of ``can some LLM agents produce behaviors $\{x_0, x_1, \dots, x_m\}$'' and are questions of generative sufficiency, rather than related to \textit{why} a behavior was produced. More generally, a lot of work uses multi-agent LLM simulations to probe the capabilities of LLMs themselves, such as their capacity for cooperation or their inherent biases. The goal of these simulations is to characterize the LLM agents' capabilities, not to necessarily explain a specific real-world social dynamic. In the survey of recent generative ABM literature by Larooij et al. \cite{larooijLargeLanguageModels2025}, we see that many projects use believability as their primary evaluation metric in this way, which we argue is an assessment of generative sufficiency rather than explanatory power.

The existence of a level 1 bucket also helps to flag if an LLM-based simulation is `cheating'. It is entirely feasible to produce LLM agents that return the outputs which produce $T$ through the manipulation of prompts. The behaviors of agents can be heavily influenced by prompt engineering; an engineered prompt that produces a desired behavior is perfectly acceptable for a level 1 phenomenal model. However, if an explanatory claim is being made, it can become important to clarify in the model's intent $I$ whether the prompt is an `artifact' that forces the correct output, or an intentional abstraction of a real-world mechanism. Without this clarification, a model may be phenomenally ambiguous.

\subsubsection{Example}
Following our social media model example, at Level 1, the researchers tweak the simulation to produce recognizable polarization patterns, operationalized as some clustering of sentiment scores over time ($T$). They audit the simulation and report that the LLM agents produce polarized discourse that human annotators rate as believable. 

At this stage, the simulation can serve two purposes. First, it demonstrates that polarization can emerge from LLM agents interacting on a simulated platform. Secondly, it could be used to forecast polarization on platforms with similar features. Given these agents on this network structure, polarization reliably emerges, and we might expect it to do so again under similar conditions. 

However, the polarization could be driven by the agents' prompts, the network topology, the recommendation algorithm, or some interaction among them. Therefore the model does not serve to identify causal responsibilities behind the phenomenon.

\subsection{Level 2}
\label{subsec:level2}
Simulations with a plausibility of level 2 move beyond reproducing $T$ to proposing a \textbf{hypothesis} for how it could possibly be generated. This is achieved when modeler specifies their Intent $I$, which includes a hypothesis and mapping (sometimes called a `model key' \cite{graebnerHowRelateModels2018}) that connects the components and activities in $S$ to the proposed mechanisms responsible for generating $T$. By doing this, one states a hypothesis about how the possible mechanisms of $T$ are related to the simulation code. Earlier, we discussed and summarized this into Kaplan and Craver's 3M requirement \cite{kaplanExplanatoryForceDynamical2011} in Section \ref{subsec:Mechanisms}.

Modeling literature often refers to these post-phenomenal simulations as `how-possibly' simulations \cite{seseljaAgentBasedModelingPhilosophy2023a, graebnerHowRelateModels2018, bokulichHowTigerBush2014} or `logical possibilities' \cite{arnoldSimulationModelsEvolution2013}. One distinguishment of level 2 from level 1 simulations is that they provide a basis for reasoning about counterfactual scenarios, given a hypothesis about $T$'s mechanisms encoded through $I$. For example, one could reason about how $T$ might change if those mechanisms were different.

Later on in Level 3, when one tries to validate a model, $I$ is what determines if the simulation behavior is right or wrong. It is worth being precise about what the mapping in $I$ involves epistemically; As discussed in Section \ref{subsec:Operationalization}, we distinguish between a model's computations and the interpretation the modeler assigns to it \cite{eganDeflatingMentalRepresentation2025a, weisbergSimulationSimilarityUsing2013}. The mapping is an interpretation that proposes how certain computational components of $S$ can be understood as standing in for certain real-world entities and activities. This means that two modelers could look at the same simulation $S$ and target $T$ and propose different mappings in $I$. 

To tie this to LLM-ABM, suppose an agent is initialized through persona prompts or steering vectors \cite{chenPersonaVectorsMonitoring2025} describing a specific profile; we can imagine the mapping in $I$ interpreting the LLM's outputs as reflecting the behavioral patterns of a `person' matching that profile. Considering the LLM's ``theory proper'' is autoregressive text prediction conditioned on a token sequence, which bears questionable structural resemblance to the cognitive processes of the described persona, we need the mapping to state: the modeler is assuming that the LLM's training data encodes the relevant distributions about human behavior. Whether this assumption holds for a given $T$ is an empirical question that is scoped to each particular domain. As this is a developing field, discerning when this assumption is reasonable is an open problem that would benefit from community discussion.

\subsubsection{Example}
At Level 2, our example researchers propose a hypothesis and mapping $I$: that polarization emerges in their simulation because agents engage with content that aligns with their initialized viewpoints, and the simulated feed algorithm amplifies this by surfacing high-engagement content. As part of the mapping, the researchers propose that the LLM agents stand in for real users on the social media platform, as the affordances made on the simulated platform (posting, replying, sharing, and receiving algorithmically ranked content) mirror the same that are available to real users. With this, future interventional questions become answerable relative to the hypothesis. For instance, ``what happens if we ablate on the recommendation algorithm?'' is now a meaningful experiment to add to evidence $E$, because the modeler has specified components they believe to be causally responsible, and exposes the hypothesis and its details to falsification. 

\begin{table}[h!]
    \centering
    \begin{tabular}{c|c|c|c|c}
        Level & $S$ & $T$ & $I$ & $E$ \\
        \hline
        0 & \checkmark & $\varnothing$ & $\varnothing$ & $\varnothing$ \\
        1 & \checkmark & \checkmark & $\varnothing$ & $\varnothing$ \\
        2 & \checkmark & \checkmark & \checkmark & $\varnothing$ \\
        3 & \checkmark & \checkmark & \checkmark & \checkmark \\
    \end{tabular}
    \caption{Plausibility levels and their relationship to the existence/falsifiability of a model's components.}
    \label{tab:PlausibilityTable}
\end{table}

\subsection{Level 3}
\label{subsec:Level3}
A simulation with plausibility level 3 attempts to ground its components in \textbf{evidence} $E$, which is used to support or constrain the model's construction, parameterization, or validation. Since the addition of each previous term in $S,T,I$ is what makes them falsifiable, this evidence could inform the design of the simulation $S$, the operationalization of the target $T$, or the justification for the mapping in $I$. $E$ could also come in the form of further constrained experiments, for example, ablations or sensitivity analyses that reinforce the prescribed hypotheses contained in the mapping. A modeler might select initial parameters based off of some observed values from census data, survey results, or prior empirical studies. For example, initializing an agent's political beliefs based on real-world polling data from a specific region might constitute a piece of evidence $E$. 

Prior work suggests that how-possibly and how-actually explanations may exist on a continuum rather than as a strict dichotomy \cite{BrandonBrandon+2014}. Metaphorically, as more evidence is gathered for the conditions postulated in an explanation, the explanation moves along the continuum until it is counted as how-actually. Adopting this viewpoint to our scale, Level 3 can be thought of not as a binary threshold like levels 0 through 2, but a gradient progressing further as the quality, quantity, and directness of that evidence increases towards an unreachable $\Omega$. We return to the question of why evidence can only asymptotically approach a definitive confirmation in Section \ref{subsec:PlausibilityClarifications}.

In our discussion in Section \ref{sec:LLMInSocialSimulation}, we elaborate on the confusion creators of LLM-based social simulations face in gathering relevant evidence $E$, particularly when much of the research effort is focused on validating the agent's internal architecture rather than the emergent social phenomenon. 

\subsubsection{Example}
Continuing the social media polarization example, at Level 3 the researchers ground the simulation by presenting varying evidence $E$. They show how agent viewpoint distributions are initialized from real survey data on political attitudes in a specific region and the recommendation algorithm mirrors a documented platform's ranking function. They run ablation studies showing that reducing the algorithmic amplification component significantly reduces polarization, consistent with prior empirical findings. Whether this evidence is appropriate is a judgment for the standards of their research domain. The model becomes more plausible as more of its components are supported, but the question of ``plausible enough'' is not one the scale answers, just makes explicit.

\begin{figure}[h!]
    \centering
    \includegraphics[width=0.7\linewidth]{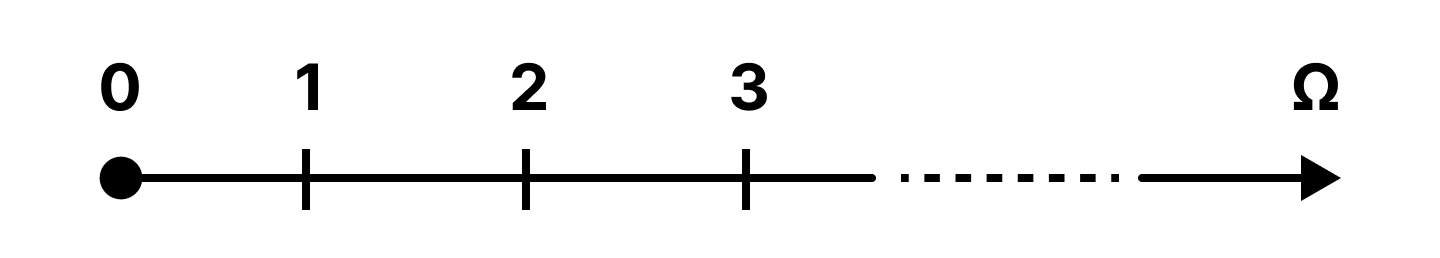}
    \caption{The Plausibility scale classifies models based on their epistemic contribution. Level $\Omega$ is considered the unreachable simulation that we can approach along level 3 continuously.}
    \label{fig:PlausibilityNumberLine}
\end{figure}

\subsection{Interpreting the Plausibility Scale} 
\label{subsec:PlausibilityClarifications}
Given these levels in their increasing order, it is not to say that simulations of a lower plausibility level are worse. It is of general agreement among scientists and philosophers that idealizations are useful, if not, necessary in building models \cite{parkerModelEvaluationAdequacyforPurpose2020, aydinonatPuzzleModelbasedExplanation2024a, elginTrueEnough2004}. Our scale clarifies the kind of epistemic contribution each simulation can provide. 
Some simulations demonstrate that a pattern can be generated, others propose and test explanations for how it can arise. Morgan and Morrison aptly describe that models can function as partially autonomous instruments that mediate between theory and data without being fully derived from either \cite{morganModelsMediatorsPerspectives1999}; they can serve as tools for exploration even when they are known to be incomplete or idealized.
For example, level 0 simulations like cellular automata can demonstrate that emergent behavior can arise from simple rules, which sets the ground for new simulation paradigms. Level 2 simulations can be used to generate ``how-possibly'' hypotheses which are falsifiable at the mechanisms level.

What increases as we move up the scale is the number of commitments the model has made that can, in principle, be shown to be wrong, and scope of the claims it can support. At Level 1, only the reproduction of $T$ is at stake. At Level 2, the mapping $I$ becomes an additional falsifiable commitment. At Level 3, the empirical grounding $E$ opens further points of potential failure. In addition, the confidence that the operationalized components of $S$ faithfully capture the abstract constructs and hypotheses the modeler intends them to represent, becomes increasingly examinable as more of the model's structure is made explicit and subject to evidence. We file all of this under the umbrella term of `plausibility'.

It is also of note that each term in $M=(S,T,I,E)$ is sequentially dependent on the previous terms for moving up plausibility levels. Consider a counterexample where a simulation $S$ only has component $E$. If there is no $T$ and $I$, the pair of $S$ and $E$ stands as a pairing of simulation outputs and arbitrary `facts', with no clear mapping between them. Thus, $T$ and $I$ are necessary relational structures that, when composed in sequence as $(S,T,I,E)$, turn sets of unrelated facts into points of evidence which support a hypothesis. This is also why Table \ref{tab:PlausibilityTable} can be helpful, as it shows each level depends on the inclusion of all previous terms.

\subsubsection{The unreachable $\Omega$ simulation}
Finally, we describe the theoretical unreachable model where every mechanism for a target phenomenon is described and leaves no doubt as to whether $S$ is a faithful representation of $T$. In the mechanist's view, to fully describe the mechanisms of a phenomenon is to explain it. We refer to this as a Level $\Omega$ simulation and note that it is a fiction we may never reach. As Brandon \cite{BrandonBrandon+2014} argues, how-possibly explanations can be thought of as a continuum toward how-actually, as more evidence is accumulated for their postulated conditions. In our scale, Level $\Omega$ (see Figure \ref{fig:PlausibilityNumberLine}) represents the (fictional) endpoint of this continuum where all postulated conditions are fully confirmed and we are sure that the mechanisms in $S$ are the mechanisms responsible for producing $T$ in actuality. However, Bokulich \cite{bokulichHowTigerBush2014} suggests that as evidence confirms a mechanism at one level of abstraction, attempts to specify that mechanism open new branches of how-possibly explanation, each requiring their own evidence. Following this, the approach toward $\Omega$ may be better thought of as a ``branching'' process in which settling one question reveals further open ones that are implicitly abstracted away when unanswered. 

Moreover, a related limitation arises from the more general relationship between evidence and theory. The Duhem-Quine thesis, loosely, holds that hypotheses are never tested in isolation, therefore the unambiguous falsification of a scientific hypothesis is impossible \cite{duhemAimStructurePhysical1954, quineLogicalPointView1953}. Another reason why we can never reach the $\Omega$ Level is that when a model is tested against empirical observations, a failure (or success) cannot unambiguously be attributed to particular components. 

It is important to characterize the $\Omega$ level both because of the inevitable idealizations introduced into models, and because it makes explicit that one cannot confirm that a simulation has fully described the mechanisms behind a phenomenon. 



\section{The Mechanism Plausibility Scale Heuristic}
\label{sec:ApplyingScale}


We draw on existing frameworks for reporting on machine learning datasets and model deployments \cite{gebruDatasheetsDatasets2021, mitchellModelCardsModel2019a, winikoffScoresheetExplainableAI2025} and present a checklist for using the Mechanism Plausibility Scale. While hypothesis testing and operationalization are long-standing, established problems in science, the novelty of LLM-ABMs may lead researchers to struggle with putting out artifacts that are epistemically cohesive, where the target phenomenon, claims, and supporting evidence are aligned and appropriately scoped to one another. In Figure \ref{fig:PlausibilityChecklist} we present the heuristic and in Appendix \ref{sec:ChecklistExamples} we follow examples using historical ABMs, one for each level in the scale.

\begin{figure}[h] 
    \centering
\begin{plausibilitybox}{Mechanism Plausibility Scale Heuristic $(S,T,I,E)$}
\label{ref:Checklist}
\small
This scale grades the model's contribution as a plausible explanation for a target phenomenon. While not every model's goal is to be explanatory, it is important to clarify when it is appropriate.
\chksection{Level 0: Simulation ($S$) -- \textit{Sandbox/Toy Model}}
\begin{itemize}[leftmargin=1.5em, label=$\square$]
    \item \textbf{Simulation.} Is the simulation (S) defined, including environments, agents, and update rules?
\end{itemize}
\noindent \textit{The requirements for Level 0 are relatively minimal; It essentially shares the same requirements for something to be considered an ABM or simulation, and mainly exists to distinguish it from the other levels. Level 0 simulations are often toys or demonstrations of new simulation paradigms (e.g., Conway's Game of Life, a demo for a new simulation framework/technique).}

\chksection{Level 1: Target Phenomenon ($T$) -- \textit{Phenomenal Model}}
\begin{itemize}[leftmargin=1.5em, label=$\square$]  
    \item \textbf{Defined Target Phenomenon.} Is a target phenomenon ($T$) operationalized (e.g. as statistical patterns, human annotations/observations, etc.)?
    \item \textbf{Generative Sufficiency.} Can the simulation ($S$) successfully generate the patterns described in ($T$)?
    \item \textbf{Reproducibility.} Does the reproducibility of the simulation (seeds, API versions, consideration of proprietary prompt injections or version changes, inherent stochasticity, etc.) match the reproducibility goals of the modeler or the field they are working in (sensitivity analysis requirements, etc.)?
\end{itemize}
\textit{The Level 1 requirement checks if there are explicit conditions that determine if the phenomena have happened and if the model produces the specified phenomena. Level 1 simulations often show that something ($T$) is possible or achievable using the entities and activities of the simulation (e.g. Can LLM agents solve games, pass theory of mind, benchmarks, etc.). }

\chksection{Level 2: Intent \& Mapping ($I$) -- \textit{How-Possibly Model}}
\begin{itemize}[leftmargin=1.5em, label=$\square$]
    \item \textbf{Simulation Contribution.} Is the simulation's use case understood (e.g., predictive, exploratory, illustrative, explanatory, etc.)?
    \item \textbf{Falsifiable Hypotheses.} Does there exist a hypothesis for how the target phenomenon $T$ arises from components in the model?
    \item \textbf{Mechanism Mapping.} Is there an Intent $I$ (implicit or explicit mapping) which connects components of the simulation ($S$) to the hypothesized `real-world' mechanisms of ($T$)?
\end{itemize}
\textit{Level 2 is a check for if the modelers have explicitly proposed how the phenomena (or parts of the phenomena) of interest are produced using the components of their simulation. This is important to make the model falsifiable.}

\chksection{Level 3: Evidence ($E$) -- \textit{Plausible Model (if validated)}}
\begin{itemize}[leftmargin=1.5em, label=$\square$]
    \item \textbf{Evidence Exists.} Is the model supported by some evidence $E$?
    \item \textbf{Relevance.} Is $E$ directed towards the claims made in $I$? Could $E$, in principle, disconfirm the hypotheses in $I$?
\end{itemize}
\textit{Adding any $E$ to get a simulation to Level 3 is straightforward; The difficulty lies in the model to remain valid, as the $E$ becomes additional falsifiable parts of the model. What is considered acceptable $E$ is bounded by the standards of the domain the model is in.}

\vspace{4pt}
\chksection{\textit{\textbf{Conclusion:} Based on the checklist, the model is classified as a \textbf{Level [X]} model.}}

\end{plausibilitybox}
\caption{The Mechanism Plausibility Scale in checklist form.}
    \label{fig:PlausibilityChecklist}
\end{figure}


\section{Discussion}\label{sec:LLMInSocialSimulation}

In the following section we review some popular ways LLMs are currently being used in simulation. Later in \ref{subsec:LLMEthics}, we engage with broader ethical and epistemic considerations for using LLM in simulation. In Section \ref{subsec:ABMIssues}, we go over historical examples and issues where the underspecification of models may have caused real-world harms. 

\subsection{Reflections on the State of LLM Social Simulation}
\label{subsec:Reflections}

``How can we use LLM social simulation practically?'' Given that no simulation can fully exhaust the mechanisms behind a phenomenon (as discussed in Section \ref{subsec:PlausibilityClarifications}) this question is best interpreted as, ``Under what conditions are LLM-ABMs adequate for a given purpose?'' 

The current state of affairs for LLM social simulation have a focus on demonstrating a simulation is capable of producing a target phenomenon. On the surface, the addition of LLMs in social simulation seemed to move us further up the ``generative sufficiency scale'', allowing agents to access a larger action space, which prompted new work in the area. This focus on generative sufficiency is reflected in the systematic review by Larooij et al., where 22 out of 35 surveyed LLM social simulation papers used `believability' as their primary validation metric \cite{larooijLargeLanguageModels2025}. Here, the believability of an agent action or simulation outcome is judged by humans or LLMs (experimentally as part of a study, or simply by inspection). A simulation validated only through believability (Level 1) may be adequate for demonstrating that a phenomenon can be generated, or for exploratory purposes such as brainstorming and prototyping. However, if a modeler wishes to test what would happen under conditions that have not been observed -- for example, how a policy intervention might alter the dynamics of $T$ -- they are implicitly making a claim about which components of $S$ are causally responsible for $T$. This is a mechanistic claim, whether or not the modeler frames it as such. 

The field of machine learning revolves around learning unknown functions or distributions from real-world observed examples. The primary goal for many papers may be predictive accuracy, and the model's internal workings are often considered a separate topic from empirical evaluations. This is no problem if prediction is the goal. However, there are a couple of caveats with LLM social simulation: many LLM-ABM projects use the evaluation of a functioning/believable LLM agent's generative sufficiency to justify the usefulness of their simulation in exploring unknown scenarios, where plausible mechanisms instead would be the relevant factor for producing relevant counterfactuals. 

We observe that LLM-based simulation is prone to conflation of agent-level validation for ABM/simulation-level validation. What do we mean by this? From the agent-based modeling perspective, a functioning agent is a presupposed mechanism -- they are generally not the target phenomena of interest. An agent's behaviors would have been manually programmed in classic agent-based models, and a non-functioning agent would have meant that the programmer made a bug. In our own attempts to review ABM papers that employ LLM-driven agents, we found that works tended to focus heavily on justifying their design of LLM-driven agents; this makes sense given that LLM-driven agents are a relatively new simulation technique. However, just as the validity of an intentional gloss is not guaranteed by the theory proper (Section \ref{subsec:Operationalization}), evidence supporting the functionality of the agent architecture (e.g., showing the agent can remember facts) is not sufficient as evidence $E$ for the mapping $I$ concerning a higher-level social phenomenon $T$. A functioning agent is a necessary part of the simulation $S$, but its functionality alone does not validate the model's explanation of $T$. To distinguish agent-level and simulation-level mechanisms, we refer to the visual metaphor in Figure \ref{fig:CraverDiagram}, which is a modified Craver diagram \cite{craverExplainingBrain2009} showing how the overall phenomena $T$ is produced by agents $\{x_1,\ldots,x_m\}$ and activities $\{\phi_1,\ldots,\phi_n\}$. 


An open question is how current LLM simulations can be made useful for policy or sociological settings given the discussed limitations so far. While we do not attempt to answer this fully, recent work suggests that practitioners already reason about simulations in ways that align with the distinctions in our scale.
Li et al. ran a year-long human co-design of simulations with their university's emergency preparedness team from 2024-2025 \cite{liWhatMakesLLM2025}. The policymakers seemed to show skepticism towards any models' predictive abilities, even if the agents exhibited believable behavior. Instead, the simulations seemed to help them more as a brainstorming tool. For example, when a simulation's dynamics were identified to be wrong, it resurfaced the policymakers' tacit knowledge and allowed them to list out important concerns, for example, wheelchair ramps in evacuation settings. This has echoes in work done by Park et al. \cite{parkSocialSimulacraCreating2022}, where `false' simulated social media platforms helped designers identify and prototype solutions to potential problems before they came up in a real deployed setting. The preparedness team began to trust the simulations more once the simulations started to align with real-world scenarios, when the authors tested it against their institution's real-world graduation commencement setting. Once the policymakers saw that the simulations generated behavior that matched outcomes based on their experience and intuition, they were willing to entertain the `how-possibly' outcomes generated by the simulation's higher-level, abstracted mechanisms.

\subsection{Proprietary LLM APIs and Reproducibility}
\label{subsec:API}

LLM API services have been known to introduce prompt injections, guardrails, or system prompts invisible to the end user. These features are added for safety, regulation, or other proprietary purposes but can be actively detrimental to experimental validity. For example, hidden prompts could unknowingly change the trajectory of an agent's behavior or prevent agents from exhibiting relevant behavior the modeler is interested in. Furthermore, proprietary LLMs are often subject to unannounced version or system prompt\footnote{See \url{https://github.com/asgeirtj/system_prompts_leaks} and similar for in-the-wild examples.} updates, which could alter agent behavior between runs. This problem is solvable with open-sourced locally hosted models, but raises the barrier to entry for many researchers because of things like GPU and technical constraints.


Concerning ethical considerations of proprietary models, LLM training data is frequently assembled through practices that fall below disciplinary ethical standards, for example, mass scraping without consent, labor practices involving underpaid workers, the inclusion of private data, and environmental harms \cite{guestCriticalArtificialIntelligence2025, weidingerEthicalSocialRisks2021}. On the methodological and epistemic side, closed-process (training sources and methods, weights) models compromise the community's ability to inspect training data, attribute model behavior to appropriate sources, and have rigorous control over their scientific methodology.

There are growing movements toward addressing these concerns. Initiatives such as AI2's OLMo project \cite{groeneveldOLMoAcceleratingScience2024} have demonstrated that competitive language models can be developed with fully open training data, code, and intermediate checkpoints, with the goal of enabling the scientific study of language models. The Public AI Network advocates for treating AI as public infrastructure -- publicly accessible, accountable, and designed to produce permanent public goods \cite{jackson_2024_13914560}. However, at the time of writing, proprietary models continue to dominate both commercial deployment and research usage.

\subsection{Broader Ethical and Epistemic Concerns}
\label{subsec:LLMEthics}
The problems related to reproducibility, proprietary APIs, and the conflation of generative sufficiency with mechanistic plausibility are largely methodological. However, there are broader ethical and epistemic concerns about the use of LLMs in social simulation that warrant consideration.

Regarding whether LLMs should serve as proxies for human subjects in the first place, Agnew et al. \cite{agnew_illusion_2024} examine proposals to substitute human research participants with LLM surrogates and find that such proposals conflict with values relating to representation, inclusion, and understanding of human subjects. Replacing participants with LLMs may disregard the relationship between researcher and subject existing in prior human subject research. When an LLM generates text that resembles survey responses or social behavior, it is not directly from the experience of a live, present individual. Furthermore, they identify the problem of ``value lock-in'', also referenced by Weidinger et al. \cite{weidingerTaxonomyRisksPosed2022}. LLMs encode the norms and attitudes present in their training data at a particular point in time. Related empirical work supports this; language models exhibit degraded performance in time periods not represented in their training corpus \cite{lazaridouMindGapAssessing2021}. 

\subsection{Historical Issues and Harms of Poor ABM specification}
\label{subsec:ABMIssues}
While the Mechanism Plausibility Scale was motivated by recent challenges posed by LLM-ABM, the ideas are not specific to ABMs with LLMs; The literature surrounding well-motivated, sound ABM design in general is a long-standing discussion \cite{swarupAdequacyWhatMakes2019, squazzoniComputationalModelsThat2020, arnoldHowModelsFail2015, northcottPrisonersDilemmaDoesnt2015, larooijLargeLanguageModels2025, vanheeLargeLanguageModels2025}. Importantly, we demonstrate how understanding a model's limits is not only important to the modeler herself, but also to its end users.

Squazzoni et al. \cite{squazzoniComputationalModelsThat2020} note that during the COVID-19 pandemic, a team at the Imperial College of London reported that results from their model projected ``a huge number of people would die in Britain unless severe policy measures were taken''. The results of their model and interventions were quickly adopted and implemented by the UK government, and advised governments of countries like the US and France in their attempts to minimize the damages caused by the virus. However, because of underspecification on what the model was adequate for, the model erroneously affected the policies of many countries, namely, being used in counterfactual scenarios when further peer analysis of the model showed it may only have been adequate for illustrative purposes. Moreover, the simulation code was not made public, even later at the time of Squazzoni et al.'s publication.

Axelrod's iterated Prisoner's Dilemma (PD) simulations are another well-known problematic case. In an adapted script on ``The Evolution of Cooperation'' \cite{axelrod_evolution_nodate}, Axelrod asserts that many real-world scenarios such as arms races, nuclear proliferation, and crisis bargaining are instances of the iterated Prisoner's Dilemma, and that advice to players of the game theoretic scenario might serve as advice to national leaders. In response, Northcott and Alexandrova \cite{northcottPrisonersDilemmaDoesnt2015} observe that despite the enormous attention devoted to the PD (over 16,000 articles since 1960), it has largely failed to explain phenomena of social scientific interest. 

Arnold (sharply) observes a broader pattern in the modeling tradition \cite{arnoldHowModelsFail2015}: over thirty years of Repeated PD simulations produced practically no successful empirical applications, yet this failure has been largely ignored. He identifies, firstly, the ``justificatory narratives'' modelers use after scrutiny, which is retreating to claims that the model is merely heuristic or exploratory without specifying the limits of that exploration. Secondly, modelers arguing that all models rely on simplification, a defense that, as Arnold notes, only holds when the causal factors a model isolates are empirically discernible from the other factors at work in the target system. When they are not, the simplification cannot be tested.

We felt it appropriate to reiterate these issues under our scale and point to related work, especially with the growing interest in simulation using LLMs. 


\section{Conclusion and Limitations}

In this paper we connect contemporary mechanisms, cognitive representation, and other philosophy of science literature with agent-based modeling and LLM social simulation. We present the Mechanism Plausibility Scale, a heuristic that classifies simulations into levels based on the falsifiability and existence of components $S,T,I,E$ and offer a practical checklist. Through a review of recent LLM-ABM papers we confirm the existence of common category errors between Agent-level and ABM-level components and underspecified models. We also connect these problems with existing issues in ABM and highlight the historical harms that occurred when these mistakes happened in high-stakes scenarios.
While our scale provides a useful heuristic, the criteria for Level 3 could be refined to differentiate the quality and extent of evidence $E$ for a more practical setting. Additionally, more could be said about a separate axis for predictive models, as opposed to our plausible explanation axis. The main focus of the paper, ultimately, remains grounding multiple disciplines in common language and bringing these issues to attention.

\section*{Generative AI Usage Statement}

This document was produced with the assistance of Generative AI, which assisted in the formatting of tables, checklists, figures, proofreading, and typographical layout of the paper. It was also used to generate critique; the authors also used AI-augmented paper search engines, such as Asta\footnote{\url{https://asta.allen.ai/chat}}, for paper discovery.





\bibliographystyle{ACM-Reference-Format}
\bibliography{facct}

@article{akataPlayingRepeatedGames2025,
	title = {Playing repeated games with {Large} {Language} {Models}},
	issn = {2397-3374},
	url = {http://arxiv.org/abs/2305.16867},
	doi = {10.1038/s41562-025-02172-y},
	urldate = {2025-05-13},
	journal = {Nature Human Behaviour},
	author = {Akata, Elif and Schulz, Lion and Coda-Forno, Julian and Oh, Seong Joon and Bethge, Matthias and Schulz, Eric},
	month = may,
	year = {2025},
	note = {arXiv:2305.16867 [cs]},
}

@misc{costarelliGameBenchEvaluatingStrategic2024,
	title = {{GameBench}: {Evaluating} {Strategic} {Reasoning} {Abilities} of {LLM} {Agents}},
	shorttitle = {{GameBench}},
	url = {http://arxiv.org/abs/2406.06613},
	doi = {10.48550/arXiv.2406.06613},
	urldate = {2025-05-13},
	publisher = {arXiv},
	author = {Costarelli, Anthony and Allen, Mat and Hauksson, Roman and Sodunke, Grace and Hariharan, Suhas and Cheng, Carlson and Li, Wenjie and Yadav, Arjun},
	month = jun,
	year = {2024},
	note = {arXiv:2406.06613 [cs]
version: 1},
}

@incollection{northcottPrisonersDilemmaDoesnt2015,
	title = {Prisoner's {Dilemma} {Doesn}'t {Explain} {Much}},
	url = {https://philarchive.org/rec/NORPDD},
	urldate = {2025-04-30},
	booktitle = {The {Prisoner}?s {Dilemma}. {Classic} philosophical arguments.},
	publisher = {Cambridge University Press},
	author = {Northcott, Robert and Alexandrova, Anna},
	editor = {Peterson, Martin},
	year = {2015},
	pages = {64--84},
}

@article{arnoldSimulationModelsEvolution2013,
	title = {Simulation {Models} of the {Evolution} of {Cooperation} as {Proofs} of {Logical} {Possibilities}. {How} {Useful} {Are} {They}?},
	volume = {15},
	url = {https://philarchive.org/rec/ARNSMO},
	number = {2},
	urldate = {2025-04-29},
	journal = {Etica E Politica},
	author = {Arnold, Eckhart},
	year = {2013},
	note = {Publisher: University of Trieste, Department of Philosophy},
	pages = {101--138},
}

@incollection{arnoldHowModelsFail2015,
	address = {Cham},
	title = {How {Models} {Fail}: {A} {Critical} {Look} at the {History} of {Computer} {Simulations} of the {Evolution} of {Cooperation}},
	isbn = {978-3-319-15514-2 978-3-319-15515-9},
	shorttitle = {How {Models} {Fail}},
	url = {https://link.springer.com/10.1007/978-3-319-15515-9_14},
	language = {en},
	urldate = {2025-04-26},
	booktitle = {Collective {Agency} and {Cooperation} in {Natural} and {Artificial} {Systems}},
	publisher = {Springer International Publishing},
	author = {Arnold, Eckhart},
	editor = {Misselhorn, Catrin},
	year = {2015},
	doi = {10.1007/978-3-319-15515-9_14},
	pages = {261--279},
}

@misc{anthisLLMSocialSimulations2025,
	title = {{LLM} {Social} {Simulations} {Are} a {Promising} {Research} {Method}},
	url = {http://arxiv.org/abs/2504.02234},
	doi = {10.48550/arXiv.2504.02234},
	urldate = {2025-04-21},
	publisher = {arXiv},
	author = {Anthis, Jacy Reese and Liu, Ryan and Richardson, Sean M. and Kozlowski, Austin C. and Koch, Bernard and Evans, James and Brynjolfsson, Erik and Bernstein, Michael},
	month = apr,
	year = {2025},
	note = {arXiv:2504.02234 [cs]},
}

@misc{parkGenerativeAgentsInteractive2023,
	title = {Generative {Agents}: {Interactive} {Simulacra} of {Human} {Behavior}},
	shorttitle = {Generative {Agents}},
	url = {http://arxiv.org/abs/2304.03442},
	doi = {10.48550/arXiv.2304.03442},
	urldate = {2025-04-21},
	publisher = {arXiv},
	author = {Park, Joon Sung and O'Brien, Joseph C. and Cai, Carrie J. and Morris, Meredith Ringel and Liang, Percy and Bernstein, Michael S.},
	month = aug,
	year = {2023},
	note = {arXiv:2304.03442 [cs]},
}

@misc{hortonLargeLanguageModels2023,
	title = {Large {Language} {Models} as {Simulated} {Economic} {Agents}: {What} {Can} {We} {Learn} from {Homo} {Silicus}?},
	shorttitle = {Large {Language} {Models} as {Simulated} {Economic} {Agents}},
	url = {http://arxiv.org/abs/2301.07543},
	doi = {10.48550/arXiv.2301.07543},
	urldate = {2025-05-22},
	publisher = {arXiv},
	author = {Horton, John J.},
	month = jan,
	year = {2023},
	note = {arXiv:2301.07543 [econ]},
}

@article{schellingModelsSegregation1969,
	title = {Models of {Segregation}},
	volume = {59},
	issn = {0002-8282},
	url = {https://www.jstor.org/stable/1823701},
	number = {2},
	urldate = {2025-05-21},
	journal = {The American Economic Review},
	author = {Schelling, Thomas C.},
	year = {1969},
	note = {Publisher: American Economic Association},
	pages = {488--493},
}

@article{craverWhenMechanisticModels2006,
	title = {When mechanistic models explain},
	volume = {153},
	issn = {1573-0964},
	url = {https://doi.org/10.1007/s11229-006-9097-x},
	doi = {10.1007/s11229-006-9097-x},
	language = {en},
	number = {3},
	urldate = {2025-08-20},
	journal = {Synthese},
	author = {Craver, Carl F.},
	month = dec,
	year = {2006},
	pages = {355--376},
}

@article{parkerModelEvaluationAdequacyforPurpose2020,
	title = {Model {Evaluation}: {An} {Adequacy}-for-{Purpose} {View}},
	volume = {87},
	copyright = {http://creativecommons.org/licenses/by/4.0/},
	issn = {0031-8248, 1539-767X},
	shorttitle = {Model {Evaluation}},
	url = {https://www.cambridge.org/core/product/identifier/S0031824800015956/type/journal_article},
	doi = {10.1086/708691},
	language = {en},
	number = {3},
	urldate = {2025-08-25},
	journal = {Philosophy of Science},
	author = {Parker, Wendy S.},
	month = jul,
	year = {2020},
	pages = {457--477},
}

@article{kaplanExplanatoryForceDynamical2011,
	title = {The {Explanatory} {Force} of {Dynamical} and {Mathematical} {Models} in {Neuroscience}: {A} {Mechanistic} {Perspective}*},
	volume = {78},
	issn = {0031-8248},
	shorttitle = {The {Explanatory} {Force} of {Dynamical} and {Mathematical} {Models} in {Neuroscience}},
	url = {https://www.jstor.org/stable/10.1086/661755},
	doi = {10.1086/661755},
	number = {4},
	urldate = {2025-08-25},
	journal = {Philosophy of Science},
	author = {Kaplan, David Michael and Craver, Carl F.},
	year = {2011},
	note = {Publisher: [The University of Chicago Press, Philosophy of Science Association]},
	pages = {601--627},
}

@incollection{craverMechanismsScience2024,
	edition = {Fall 2024},
	title = {Mechanisms in {Science}},
	url = {https://plato.stanford.edu/archives/fall2024/entries/science-mechanisms/},
	urldate = {2025-08-25},
	booktitle = {The {Stanford} {Encyclopedia} of {Philosophy}},
	publisher = {Metaphysics Research Lab, Stanford University},
	author = {Craver, Carl and Tabery, James and Illari, Phyllis},
	editor = {Zalta, Edward N. and Nodelman, Uri},
	year = {2024},
}

@incollection{aydinonatPuzzleModelbasedExplanation2024a,
	address = {London},
	edition = {1},
	title = {The puzzle of model-based explanation},
	isbn = {978-1-003-20564-7},
	url = {https://www.taylorfrancis.com/books/9781003205647/chapters/10.4324/9781003205647-16},
	language = {en},
	urldate = {2025-08-27},
	booktitle = {The {Routledge} {Handbook} of {Philosophy} of {Scientific} {Modeling}},
	publisher = {Routledge},
	author = {Aydinonat, N. Emrah},
	collaborator = {Knuuttila, Tarja and Carrillo, Natalia and Koskinen, Rami},
	month = aug,
	year = {2024},
	doi = {10.4324/9781003205647-16},
	pages = {177--192},
}

@article{graebnerHowRelateModels2018,
	title = {How to {Relate} {Models} to {Reality}? {An} {Epistemological} {Framework} for the {Validation} and {Verification} of {Computational} {Models}},
	volume = {21},
	issn = {1460-7425},
	shorttitle = {How to {Relate} {Models} to {Reality}?},
	number = {3},
	journal = {Journal of Artificial Societies and Social Simulation},
	author = {Graebner, Claudius},
	year = {2018},
	pages = {8},
}

@article{weisbergWhoModeler2007,
	title = {Who {Is} a {Modeler}?},
	volume = {58},
	issn = {0007-0882},
	url = {https://www.jstor.org/stable/30115224},
	number = {2},
	urldate = {2025-08-29},
	journal = {The British Journal for the Philosophy of Science},
	author = {Weisberg, Michael},
	year = {2007},
	note = {Publisher: [Oxford University Press, The British Society for the Philosophy of Science]},
	pages = {207--233},
}

@book{weisbergSimulationSimilarityUsing2013,
	title = {Simulation and {Similarity}: {Using} {Models} to {Understand} the {World}},
	publisher = {Oxford University Press},
	author = {Weisberg, Michael},
	year = {2013},
}

@book{epsteinGenerativeSocialScience2006,
	edition = {STU - Student edition},
	title = {Generative {Social} {Science}: {Studies} in {Agent}-{Based} {Computational} {Modeling}},
	isbn = {978-0-691-12547-3},
	url = {http://www.jstor.org/stable/j.ctt7rxj1},
	urldate = {2025-09-09},
	publisher = {Princeton University Press},
	author = {Epstein, Joshua M.},
	year = {2006},
}

@incollection{seseljaAgentBasedModelingPhilosophy2023a,
	edition = {Winter 2023},
	title = {Agent-{Based} {Modeling} in the {Philosophy} of {Science}},
	url = {https://plato.stanford.edu/archives/win2023/entries/agent-modeling-philscience/},
	urldate = {2025-09-12},
	booktitle = {The {Stanford} {Encyclopedia} of {Philosophy}},
	publisher = {Metaphysics Research Lab, Stanford University},
	author = {Šešelja, Dunja},
	editor = {Zalta, Edward N. and Nodelman, Uri},
	year = {2023},
}

@article{maukPotentialEffectivenessSimulations2000,
	title = {The potential effectiveness of simulations versus phenomenological models},
	volume = {3},
	copyright = {2000 Nature America Inc.},
	issn = {1546-1726},
	url = {https://www.nature.com/articles/nn0700_649},
	doi = {10.1038/76606},
	language = {en},
	number = {7},
	urldate = {2025-09-12},
	journal = {Nature Neuroscience},
	author = {Mauk, Michael D.},
	month = jul,
	year = {2000},
	note = {Publisher: Nature Publishing Group},
	pages = {649--651},
}

@article{kayPrinciplesModelsNeural2018,
	series = {New advances in encoding and decoding of brain signals},
	title = {Principles for models of neural information processing},
	volume = {180},
	issn = {1053-8119},
	url = {https://www.sciencedirect.com/science/article/pii/S1053811917306638},
	doi = {10.1016/j.neuroimage.2017.08.016},
	urldate = {2025-09-12},
	journal = {NeuroImage},
	author = {Kay, Kendrick N.},
	month = oct,
	year = {2018},
	pages = {101--109},
}

@book{fisherGeneticalTheoryNatural1999,
	address = {Oxford},
	edition = {A complete variorum ed},
	title = {The genetical theory of natural selection: by {R}.{A}. {Fisher} ; edited with a foreword and notes by {J}.{H}. {Bennett}},
	isbn = {978-0-19-850440-5},
	shorttitle = {The genetical theory of natural selection},
	publisher = {Oxford University Press},
	author = {Fisher, Ronald Aylmer},
	year = {1999},
}

@misc{guoGPTGameTheory2023,
	title = {{GPT} in {Game} {Theory} {Experiments}},
	url = {http://arxiv.org/abs/2305.05516},
	doi = {10.48550/arXiv.2305.05516},
	urldate = {2025-09-16},
	publisher = {arXiv},
	author = {Guo, Fulin},
	month = dec,
	year = {2023},
	note = {arXiv:2305.05516 [econ]},
}

@article{parkSocialSimulacraCreating2022,
	title = {Social {Simulacra}: {Creating} {Populated} {Prototypes} for {Social} {Computing} {Systems}},
	shorttitle = {Social {Simulacra}},
	url = {https://dl.acm.org/doi/10.1145/3526113.3545616},
	doi = {10.1145/3526113.3545616},
	language = {en},
	urldate = {2025-09-16},
	journal = {Proceedings of the 35th Annual ACM Symposium on User Interface Software and Technology},
	author = {Park, Joon Sung and Popowski, Lindsay and Cai, Carrie and Morris, Meredith Ringel and Liang, Percy and Bernstein, Michael S.},
	month = oct,
	year = {2022},
	note = {Conference Name: UIST '22: The 35th Annual ACM Symposium on User Interface Software and Technology
ISBN: 9781450393201
Place: Bend OR USA
Publisher: ACM},
	pages = {1--18},
}

@misc{huaWarPeaceWarAgent2024,
	title = {War and {Peace} ({WarAgent}): {Large} {Language} {Model}-based {Multi}-{Agent} {Simulation} of {World} {Wars}},
	shorttitle = {War and {Peace} ({WarAgent})},
	url = {http://arxiv.org/abs/2311.17227},
	doi = {10.48550/arXiv.2311.17227},
	urldate = {2025-09-16},
	publisher = {arXiv},
	author = {Hua, Wenyue and Fan, Lizhou and Li, Lingyao and Mei, Kai and Ji, Jianchao and Ge, Yingqiang and Hemphill, Libby and Zhang, Yongfeng},
	month = jan,
	year = {2024},
	note = {arXiv:2311.17227 [cs]},
}

@misc{larooijLargeLanguageModels2025,
	title = {Do {Large} {Language} {Models} {Solve} the {Problems} of {Agent}-{Based} {Modeling}? {A} {Critical} {Review} of {Generative} {Social} {Simulations}},
	shorttitle = {Do {Large} {Language} {Models} {Solve} the {Problems} of {Agent}-{Based} {Modeling}?},
	url = {http://arxiv.org/abs/2504.03274},
	doi = {10.48550/arXiv.2504.03274},
	urldate = {2025-09-17},
	publisher = {arXiv},
	author = {Larooij, Maik and Törnberg, Petter},
	month = apr,
	year = {2025},
	note = {arXiv:2504.03274 [cs]},
}

@article{machamerThinkingMechanisms2000,
	title = {Thinking about {Mechanisms}},
	volume = {67},
	issn = {0031-8248},
	url = {https://www.jstor.org/stable/188611},
	number = {1},
	urldate = {2025-09-18},
	journal = {Philosophy of Science},
	author = {Machamer, Peter and Darden, Lindley and Craver, Carl F.},
	year = {2000},
	note = {Publisher: [The University of Chicago Press, Philosophy of Science Association]},
	pages = {1--25},
}

@book{glennanNewMechanicalPhilosophy2017,
	address = {Oxford},
	title = {The {New} {Mechanical} {Philosophy}},
	publisher = {Oxford University Press},
	author = {Glennan, Stuart},
	year = {2017},
}

@misc{kaiyaLyfeAgentsGenerative2023,
  title = {Lyfe {{Agents}}: {{Generative}} Agents for Low-Cost Real-Time Social Interactions},
  shorttitle = {Lyfe {{Agents}}},
  author = {Kaiya, Zhao and Naim, Michelangelo and Kondic, Jovana and Cortes, Manuel and Ge, Jiaxin and Luo, Shuying and Yang, Guangyu Robert and Ahn, Andrew},
  year = {2023},
  month = oct,
  number = {arXiv:2310.02172},
  eprint = {2310.02172},
  primaryclass = {cs},
  publisher = {arXiv},
  doi = {10.48550/arXiv.2310.02172},
  urldate = {2025-09-26},
  archiveprefix = {arXiv},
}

@book{carminesReliabilityValidityAssessment1979,
  title = {Reliability and {{Validity Assessment}}},
  author = {G.Carmines, Edward and A.Zeller, Richard},
  year = {1979},
  publisher = {SAGE Publications, Inc.},
  doi = {10.4135/9781412985642},
  urldate = {2025-10-04},
  isbn = {978-1-4129-8564-2},
  langid = {english}
}

@article{stevensOperationalDefinitionPsychological1935,
  title = {The Operational Definition of Psychological Concepts},
  author = {Stevens, S. S.},
  year = {1935},
  journal = {Psychological Review},
  volume = {42},
  number = {6},
  pages = {517--527},
  publisher = {Psychological Review Company},
  address = {US},
  issn = {1939-1471},
  doi = {10.1037/h0056973}
}

@book{bridgmanLogicModernPhysics1927,
  title = {The Logic of Modern Physics},
  author = {Bridgman, P. W. (Percy Williams)},
  year = {1927},
  publisher = {The Macmillan Company},
  urldate = {2025-10-04},
  copyright = {Public domain in the USA.},
  langid = {english},
  lccn = {EBook-No. 70620}
}

@article{vessonenConceptualEngineeringOperationalism2021,
	title = {Conceptual engineering and operationalism in psychology},
	volume = {199},
	issn = {1573-0964},
	url = {https://doi.org/10.1007/s11229-021-03261-x},
	doi = {10.1007/s11229-021-03261-x},
	language = {en},
	number = {3},
	urldate = {2025-10-04},
	journal = {Synthese},
	author = {Vessonen, Elina},
	month = dec,
	year = {2021},
	pages = {10615--10637},
}

@article{cronbachConstructValidityPsychological1955,
	title = {Construct validity in psychological tests},
	volume = {52},
	issn = {1939-1455},
	doi = {10.1037/h0040957},
	number = {4},
	journal = {Psychological Bulletin},
	author = {Cronbach, Lee J. and Meehl, Paul E.},
	year = {1955},
	note = {Place: US
Publisher: American Psychological Association},
	pages = {281--302},
}

@article{glennanMechanismsNatureCausation1996,
	title = {Mechanisms and the nature of causation},
	volume = {44},
	issn = {1572-8420},
	url = {https://doi.org/10.1007/BF00172853},
	doi = {10.1007/BF00172853},
	language = {en},
	number = {1},
	urldate = {2025-10-05},
	journal = {Erkenntnis},
	author = {Glennan, Stuart S.},
	month = jan,
	year = {1996},
	pages = {49--71},
}

@article{pichlerConceptsTextsBack2022,
	title = {From {Concepts} to {Texts} and {Back}: {Operationalization} as a {Core} {Activity} of {Digital} {Humanities}},
	volume = {7},
	copyright = {http://creativecommons.org/licenses/by/4.0},
	issn = {2371-4549},
	shorttitle = {From {Concepts} to {Texts} and {Back}},
	url = {https://culturalanalytics.org/article/57195-from-concepts-to-texts-and-back-operationalization-as-a-core-activity-of-digital-humanities},
	doi = {10.22148/001c.57195},
	language = {en},
	number = {4},
	urldate = {2025-10-04},
	journal = {Journal of Cultural Analytics},
	author = {Pichler, Axel and Reiter, Nils},
	month = dec,
	year = {2022},
}

@book{craverExplainingBrain2009,
	title = {Explaining the {Brain}},
	publisher = {Oxford University Press},
	author = {Craver, Carl F.},
	year = {2009},
}

@online{gaoS3SocialnetworkSimulation2025,
  title = {S\$\textasciicircum 3\$: {{Social-network Simulation System}} with {{Large Language Model-Empowered Agents}}},
  shorttitle = {S\$\textasciicircum 3\$},
  author = {Gao, Chen and Lan, Xiaochong and Lu, Zhihong and Mao, Jinzhu and Piao, Jinghua and Wang, Huandong and Jin, Depeng and Li, Yong},
  date = {2025-06-03},
  eprint = {2307.14984},
  eprinttype = {arXiv},
  eprintclass = {cs},
  doi = {10.48550/arXiv.2307.14984},
  url = {http://arxiv.org/abs/2307.14984},
  urldate = {2025-10-06},
  pubstate = {prepublished}
}

@misc{wangHumanoidAgentsPlatform2023,
	title = {Humanoid {Agents}: {Platform} for {Simulating} {Human}-like {Generative} {Agents}},
	shorttitle = {Humanoid {Agents}},
	url = {http://arxiv.org/abs/2310.05418},
	doi = {10.48550/arXiv.2310.05418},
	language = {en},
	urldate = {2025-09-26},
	publisher = {arXiv},
	author = {Wang, Zhilin and Chiu, Yu Ying and Chiu, Yu Cheung},
	month = oct,
	year = {2023},
	note = {arXiv:2310.05418 [cs]},
}

@misc{chenAgentVerseFacilitatingMultiAgent2023,
	title = {{AgentVerse}: {Facilitating} {Multi}-{Agent} {Collaboration} and {Exploring} {Emergent} {Behaviors}},
	shorttitle = {{AgentVerse}},
	url = {http://arxiv.org/abs/2308.10848},
	doi = {10.48550/arXiv.2308.10848},
	urldate = {2025-10-06},
	publisher = {arXiv},
	author = {Chen, Weize and Su, Yusheng and Zuo, Jingwei and Yang, Cheng and Yuan, Chenfei and Chan, Chi-Min and Yu, Heyang and Lu, Yaxi and Hung, Yi-Hsin and Qian, Chen and Qin, Yujia and Cong, Xin and Xie, Ruobing and Liu, Zhiyuan and Sun, Maosong and Zhou, Jie},
	month = oct,
	year = {2023},
	note = {arXiv:2308.10848 [cs]},
}

@misc{williamsEpidemicModelingGenerative2023,
	title = {Epidemic {Modeling} with {Generative} {Agents}},
	url = {http://arxiv.org/abs/2307.04986},
	doi = {10.48550/arXiv.2307.04986},
	urldate = {2025-10-06},
	publisher = {arXiv},
	author = {Williams, Ross and Hosseinichimeh, Niyousha and Majumdar, Aritra and Ghaffarzadegan, Navid},
	month = jul,
	year = {2023},
	note = {arXiv:2307.04986 [cs]},
}

@misc{alProjectSidManyagent2024,
	title = {Project {Sid}: {Many}-agent simulations toward {AI} civilization},
	shorttitle = {Project {Sid}},
	url = {http://arxiv.org/abs/2411.00114},
	doi = {10.48550/arXiv.2411.00114},
	urldate = {2025-10-06},
	publisher = {arXiv},
	author = {AL, Altera and Ahn, Andrew and Becker, Nic and Carroll, Stephanie and Christie, Nico and Cortes, Manuel and Demirci, Arda and Du, Melissa and Li, Frankie and Luo, Shuying and Wang, Peter Y. and Willows, Mathew and Yang, Feitong and Yang, Guangyu Robert},
	month = oct,
	year = {2024},
	note = {arXiv:2411.00114 [cs]},
}

@misc{liLargeLanguageModeldriven2024,
	title = {Large {Language} {Model}-driven {Multi}-{Agent} {Simulation} for {News} {Diffusion} {Under} {Different} {Network} {Structures}},
	url = {http://arxiv.org/abs/2410.13909},
	doi = {10.48550/arXiv.2410.13909},
	urldate = {2025-10-06},
	publisher = {arXiv},
	author = {Li, Xinyi and Xu, Yu and Zhang, Yongfeng and Malthouse, Edward C.},
	month = oct,
	year = {2024},
	note = {arXiv:2410.13909 [cs]},
}

@misc{zhangSpeechAgentsHumanCommunicationSimulation2024,
	title = {{SpeechAgents}: {Human}-{Communication} {Simulation} with {Multi}-{Modal} {Multi}-{Agent} {Systems}},
	shorttitle = {{SpeechAgents}},
	url = {http://arxiv.org/abs/2401.03945},
	doi = {10.48550/arXiv.2401.03945},
	urldate = {2025-10-06},
	publisher = {arXiv},
	author = {Zhang, Dong and Li, Zhaowei and Wang, Pengyu and Zhang, Xin and Zhou, Yaqian and Qiu, Xipeng},
	month = jan,
	year = {2024},
	note = {arXiv:2401.03945 [cs]},
}

@inproceedings{liuSkepticismAcceptanceSimulating2024,
	title = {From {Skepticism} to {Acceptance}: {Simulating} the {Attitude} {Dynamics} {Toward} {Fake} {News}},
	shorttitle = {From {Skepticism} to {Acceptance}},
	url = {http://arxiv.org/abs/2403.09498},
	doi = {10.24963/ijcai.2024/873},
	urldate = {2025-10-06},
	booktitle = {Proceedings of the {Thirty}-{ThirdInternational} {Joint} {Conference} on {Artificial} {Intelligence}},
	author = {Liu, Yuhan and Chen, Xiuying and Zhang, Xiaoqing and Gao, Xing and Zhang, Ji and Yan, Rui},
	month = aug,
	year = {2024},
	note = {arXiv:2403.09498 [cs]},
	pages = {7849--7857},
}

@misc{marzoEmergenceScaleFreeNetworks2023,
	title = {Emergence of {Scale}-{Free} {Networks} in {Social} {Interactions} among {Large} {Language} {Models}},
	url = {http://arxiv.org/abs/2312.06619},
	doi = {10.48550/arXiv.2312.06619},
	urldate = {2025-09-26},
	publisher = {arXiv},
	author = {Marzo, Giordano De and Pietronero, Luciano and Garcia, David},
	month = dec,
	year = {2023},
	note = {arXiv:2312.06619 [physics]},
}

@misc{zhangExploringCollaborationMechanisms2024,
	title = {Exploring {Collaboration} {Mechanisms} for {LLM} {Agents}: {A} {Social} {Psychology} {View}},
	shorttitle = {Exploring {Collaboration} {Mechanisms} for {LLM} {Agents}},
	url = {http://arxiv.org/abs/2310.02124},
	doi = {10.48550/arXiv.2310.02124},
	urldate = {2025-10-06},
	publisher = {arXiv},
	author = {Zhang, Jintian and Xu, Xin and Zhang, Ningyu and Liu, Ruibo and Hooi, Bryan and Deng, Shumin},
	month = may,
	year = {2024},
	note = {arXiv:2310.02124 [cs]},
}

@inproceedings{liTheoryMindMultiAgent2023,
	title = {Theory of {Mind} for {Multi}-{Agent} {Collaboration} via {Large} {Language} {Models}},
	url = {http://arxiv.org/abs/2310.10701},
	doi = {10.18653/v1/2023.emnlp-main.13},
	urldate = {2025-10-06},
	booktitle = {Proceedings of the 2023 {Conference} on {Empirical} {Methods} in {Natural} {Language} {Processing}},
	author = {Li, Huao and Chong, Yu Quan and Stepputtis, Simon and Campbell, Joseph and Hughes, Dana and Lewis, Michael and Sycara, Katia},
	year = {2023},
	note = {arXiv:2310.10701 [cs]},
	pages = {180--192},
}

@article{elginTrueEnough2004,
	title = {True {Enough}},
	volume = {14},
	issn = {1533-6077},
	url = {https://www.jstor.org/stable/3050623},
	urldate = {2025-10-07},
	journal = {Philosophical Issues},
	author = {Elgin, Catherine Z.},
	year = {2004},
	note = {Publisher: [Wiley, Ridgeview Publishing Company]},
	pages = {113--131},
}

@article{gardnerMathematicalGames1970,
	title = {Mathematical {Games}},
	volume = {223},
	issn = {0036-8733},
	url = {https://www.jstor.org/stable/24927642},
	number = {4},
	urldate = {2026-01-12},
	journal = {Scientific American},
	author = {Gardner, Martin},
	year = {1970},
	note = {Publisher: Scientific American, a division of Nature America, Inc.},
	pages = {120--123},
}

@article{landisMeasurementObserverAgreement1977,
	title = {The measurement of observer agreement for categorical data},
	volume = {33},
	issn = {0006-341X},
	language = {eng},
	number = {1},
	journal = {Biometrics},
	author = {Landis, J. R. and Koch, G. G.},
	month = mar,
	year = {1977},
	pmid = {843571},
	pages = {159--174},
}

@misc{winikoffScoresheetExplainableAI2025,
	title = {A {Scoresheet} for {Explainable} {AI}},
	url = {http://arxiv.org/abs/2502.09861},
	doi = {10.48550/arXiv.2502.09861},
	urldate = {2025-09-30},
	publisher = {arXiv},
	author = {Winikoff, Michael and Thangarajah, John and Rodriguez, Sebastian},
	month = feb,
	year = {2025},
	note = {arXiv:2502.09861 [cs]},
}

@inproceedings{kempinskiGameThoughtsIterative2025,
	address = {Richland, SC},
	series = {{AAMAS} '25},
	title = {Game of {Thoughts}: {Iterative} {Reasoning} in {Game}-{Theoretic} {Domains} with {Large} {Language} {Models}},
	isbn = {979-8-4007-1426-9},
	shorttitle = {Game of {Thoughts}},
	urldate = {2025-10-08},
	booktitle = {Proceedings of the 24th {International} {Conference} on {Autonomous} {Agents} and {Multiagent} {Systems}},
	publisher = {International Foundation for Autonomous Agents and Multiagent Systems},
	author = {Kempinski, Benjamin and Gemp, Ian and Larson, Kate and Lanctot, Marc and Bachrach, Yoram and Kachman, Tal},
	month = jun,
	year = {2025},
	pages = {1088--1097},
}

@misc{gebruDatasheetsDatasets2021,
	title = {Datasheets for {Datasets}},
	url = {http://arxiv.org/abs/1803.09010},
	doi = {10.48550/arXiv.1803.09010},
	urldate = {2025-10-09},
	publisher = {arXiv},
	author = {Gebru, Timnit and Morgenstern, Jamie and Vecchione, Briana and Vaughan, Jennifer Wortman and Wallach, Hanna and III, Hal Daumé and Crawford, Kate},
	month = dec,
	year = {2021},
	note = {arXiv:1803.09010 [cs]},
}

@article{wangUserBehaviorSimulation2025,
	title = {User {Behavior} {Simulation} with {Large} {Language} {Model}-based {Agents}},
	volume = {43},
	issn = {1046-8188},
	url = {https://dl.acm.org/doi/10.1145/3708985},
	doi = {10.1145/3708985},
	number = {2},
	urldate = {2025-10-09},
	journal = {ACM Trans. Inf. Syst.},
	author = {Wang, Lei and Zhang, Jingsen and Yang, Hao and Chen, Zhi-Yuan and Tang, Jiakai and Zhang, Zeyu and Chen, Xu and Lin, Yankai and Sun, Hao and Song, Ruihua and Zhao, Xin and Xu, Jun and Dou, Zhicheng and Wang, Jun and Wen, Ji-Rong},
	month = jan,
	year = {2025},
	pages = {55:1--55:37},
}

@misc{renEmergenceSocialNorms2024,
	title = {Emergence of {Social} {Norms} in {Generative} {Agent} {Societies}: {Principles} and {Architecture}},
	shorttitle = {Emergence of {Social} {Norms} in {Generative} {Agent} {Societies}},
	url = {http://arxiv.org/abs/2403.08251},
	doi = {10.48550/arXiv.2403.08251},
	urldate = {2025-10-09},
	publisher = {arXiv},
	author = {Ren, Siyue and Cui, Zhiyao and Song, Ruiqi and Wang, Zhen and Hu, Shuyue},
	month = aug,
	year = {2024},
	note = {arXiv:2403.08251 [cs]},
}

@inproceedings{paulMakingReasoningMatter2024,
	title = {Making {Reasoning} {Matter}: {Measuring} and {Improving} {Faithfulness} of {Chain}-of-{Thought} {Reasoning}},
	copyright = {Creative Commons Attribution 4.0 International},
	shorttitle = {Making {Reasoning} {Matter}},
	url = {https://arxiv.org/abs/2402.13950},
	doi = {10.48550/ARXIV.2402.13950},
	urldate = {2025-10-09},
	publisher = {arXiv},
	author = {Paul, Debjit and West, Robert and Bosselut, Antoine and Faltings, Boi},
	year = {2024},
	note = {Version Number: 4},
}

@article{yonaCanLargeLanguage2024,
	title = {Can {Large} {Language} {Models} {Faithfully} {Express} {Their} {Intrinsic} {Uncertainty} in {Words}?},
	copyright = {Creative Commons Attribution 4.0 International},
	url = {https://arxiv.org/abs/2405.16908},
	doi = {10.48550/ARXIV.2405.16908},
	urldate = {2025-10-09},
	author = {Yona, Gal and Aharoni, Roee and Geva, Mor},
	year = {2024},
	note = {Publisher: arXiv
Version Number: 2},
}

@misc{li_spontaneous_2025,
	title = {Spontaneous {Giving} and {Calculated} {Greed} in {Language} {Models}},
	url = {http://arxiv.org/abs/2502.17720},
	doi = {10.48550/arXiv.2502.17720},
	urldate = {2025-05-16},
	publisher = {arXiv},
	author = {Li, Yuxuan and Shirado, Hirokazu},
	month = mar,
	year = {2025},
	note = {arXiv:2502.17720 [cs]},
}

@incollection{bartha_analogy_2024,
	edition = {Fall 2024},
	title = {Analogy and {Analogical} {Reasoning}},
	url = {https://plato.stanford.edu/archives/fall2024/entries/reasoning-analogy/},
	urldate = {2026-01-06},
	booktitle = {The {Stanford} {Encyclopedia} of {Philosophy}},
	publisher = {Metaphysics Research Lab, Stanford University},
	author = {Bartha, Paul},
	editor = {Zalta, Edward N. and Nodelman, Uri},
	year = {2024},
}

@inproceedings{mitchellModelCardsModel2019a,
	title = {Model {Cards} for {Model} {Reporting}},
	url = {http://arxiv.org/abs/1810.03993},
	doi = {10.1145/3287560.3287596},
	urldate = {2026-01-06},
	booktitle = {Proceedings of the {Conference} on {Fairness}, {Accountability}, and {Transparency}},
	author = {Mitchell, Margaret and Wu, Simone and Zaldivar, Andrew and Barnes, Parker and Vasserman, Lucy and Hutchinson, Ben and Spitzer, Elena and Raji, Inioluwa Deborah and Gebru, Timnit},
	month = jan,
	year = {2019},
	note = {arXiv:1810.03993 [cs]},
	pages = {220--229},
}

@misc{liWhatMakesLLM2025,
	title = {What {Makes} {LLM} {Agent} {Simulations} {Useful} for {Policy}? {Insights} {From} an {Iterative} {Design} {Engagement} in {Emergency} {Preparedness}},
	shorttitle = {What {Makes} {LLM} {Agent} {Simulations} {Useful} for {Policy}?},
	url = {http://arxiv.org/abs/2509.21868},
	doi = {10.48550/arXiv.2509.21868},
	urldate = {2026-01-11},
	publisher = {arXiv},
	author = {Li, Yuxuan and Das, Sauvik and Shirado, Hirokazu},
	month = sep,
	year = {2025},
	note = {arXiv:2509.21868 [cs]},
}

@misc{vanheeLargeLanguageModels2025,
	title = {Large {Language} {Models} for {Agent}-{Based} {Modelling}: {Current} and possible uses across the modelling cycle},
	shorttitle = {Large {Language} {Models} for {Agent}-{Based} {Modelling}},
	url = {http://arxiv.org/abs/2507.05723},
	doi = {10.48550/arXiv.2507.05723},
	urldate = {2025-11-21},
	publisher = {arXiv},
	author = {Vanhée, Loïs and Borit, Melania and Siebers, Peer-Olaf and Cremades, Roger and Frantz, Christopher and Gürcan, Önder and Kalvas, František and Kera, Denisa Reshef and Nallur, Vivek and Narasimhan, Kavin and Neumann, Martin},
	month = jul,
	year = {2025},
	note = {arXiv:2507.05723 [cs]
version: 1},
}

@inproceedings{swarupAdequacyWhatMakes2019,
	title = {Adequacy: {What} {Makes} a {Simulation} {Good} {Enough}?},
	shorttitle = {Adequacy},
	url = {https://ieeexplore.ieee.org/abstract/document/8732895/authors},
	doi = {10.23919/SpringSim.2019.8732895},
	urldate = {2025-11-21},
	booktitle = {2019 {Spring} {Simulation} {Conference} ({SpringSim})},
	author = {Swarup, Samarth},
	month = apr,
	year = {2019},
	pages = {1--12},
}

@article{squazzoniComputationalModelsThat2020,
	title = {Computational {Models} {That} {Matter} {During} a {Global} {Pandemic} {Outbreak}: {A} {Call} to {Action}},
	volume = {23},
	issn = {1460-7425},
	shorttitle = {Computational {Models} {That} {Matter} {During} a {Global} {Pandemic} {Outbreak}},
	doi = {10.18564/jasss.4298},
	number = {2},
	journal = {JASSS - The Journal of Artificial Societies and Social Simulation},
	author = {Squazzoni, Flaminio and Polhill, J. Gareth and Edmonds, Bruce and Ahrweiler, Petra and Antosz, Patrycja and Scholz, Geeske and Chappin, Emile and Borit, Melania and Verhagen, Harko and Giardini, Francesca and Gilbert, Nigel},
	month = mar,
	year = {2020},
}

@article{bogenSavingPhenomena1988,
  title = {Saving the {{Phenomena}}},
  author = {Bogen, James and Woodward, James},
  year = 1988,
  journal = {The Philosophical Review},
  volume = {97},
  number = {3},
  eprint = {2185445},
  eprinttype = {jstor},
  pages = {303--352},
  publisher = {[Duke University Press, Philosophical Review]},
  issn = {0031-8108},
  doi = {10.2307/2185445},
  urldate = {2026-01-13}
}

@article{massimiPerspectivalOntologySituated2022,
  title = {Perspectival {{Ontology}}: {{Between Situated Knowledge}} and {{Multiculturalism}}},
  shorttitle = {Perspectival {{Ontology}}},
  author = {Massimi, Michela},
  year = 2022,
  month = mar,
  journal = {The Monist},
  volume = {105},
  number = {2},
  pages = {214--228},
  issn = {0026-9662, 2153-3601},
  doi = {10.1093/monist/onab032},
  urldate = {2026-01-14},
  copyright = {https://academic.oup.com/journals/pages/open\_access/funder\_policies/chorus/standard\_publication\_model},
  langid = {english}
}

@article{mcallisterPhenomenaPatternsData1997,
  title = {Phenomena and {{Patterns}} in {{Data Sets}}},
  author = {McAllister, James W.},
  year = 1997,
  journal = {Erkenntnis (1975-)},
  volume = {47},
  number = {2},
  eprint = {20012798},
  eprinttype = {jstor},
  pages = {217--228},
  publisher = {Springer},
  issn = {0165-0106},
  doi = {10.1023/A:1005387021520},
  urldate = {2026-01-13}
}

@article{axelrod_evolution_nodate,
	title = {The {Evolution} of {Cooperation}*},
	url = {https://ee.stanford.edu/~hellman/Breakthrough/book/pdfs/axelrod.pdf},
	language = {en},
	author = {Axelrod, Robert},
}

@inproceedings{agnew_illusion_2024,
	title = {The illusion of artificial inclusion},
	url = {http://arxiv.org/abs/2401.08572},
	doi = {10.1145/3613904.3642703},
	urldate = {2025-04-28},
	booktitle = {Proceedings of the {CHI} {Conference} on {Human} {Factors} in {Computing} {Systems}},
	author = {Agnew, William and Bergman, A. Stevie and Chien, Jennifer and Díaz, Mark and El-Sayed, Seliem and Pittman, Jaylen and Mohamed, Shakir and McKee, Kevin R.},
	month = may,
	year = {2024},
	note = {arXiv:2401.08572 [cs]},
	pages = {1--12},
}

@article{maccorquodaleDistinctionHypotheticalConstructs1948,
  title = {On a Distinction between Hypothetical Constructs and Intervening Variables.},
  author = {MacCorquodale, Kenneth and Meehl, Paul E.},
  year = 1948,
  journal = {Psychological Review},
  volume = {55},
  number = {2},
  pages = {95--107},
  issn = {1939-1471, 0033-295X},
  doi = {10.1037/h0056029},
  urldate = {2026-01-14},
  langid = {english}
}

@article{jacksonEpiphenomenalQualia1982,
  title = {Epiphenomenal {{Qualia}}},
  author = {Jackson, Frank},
  year = 1982,
  month = apr,
  journal = {The Philosophical Quarterly},
  volume = {32},
  number = {127},
  pages = {127--136},
  issn = {0031-8094},
  doi = {10.2307/2960077},
  urldate = {2026-01-14}
}

@book{titchenerTextbookPsychology1910,
  title = {A Text-Book of Psychology},
  author = {Titchener, Edward Bradford},
  year = 1910,
  series = {A Text-Book of Psychology},
  pages = {xx, 565},
  publisher = {MacMillan Co},
  address = {New York, NY, US},
  doi = {10.1037/10907-000}
}

@book{pearlCausality2009,
	address = {Cambridge},
	edition = {2},
	title = {Causality},
	isbn = {978-0-521-89560-6},
	url = {https://www.cambridge.org/core/books/causality/B0046844FAE10CBF274D4ACBDAEB5F5B},
	doi = {10.1017/CBO9780511803161},
	urldate = {2026-03-19},
	publisher = {Cambridge University Press},
	author = {Pearl, Judea},
	year = {2009},
}

@book{10.5555/3238230,
	address = {USA},
	edition = {1},
	title = {The book of why: {The} new science of cause and effect},
	isbn = {0-465-09760-X},
	publisher = {Basic Books, Inc.},
	author = {Pearl, Judea and Mackenzie, Dana},
	year = {2018},
}

@article{bokulichHowTigerBush2014,
	title = {How the {Tiger} {Bush} {Got} its {Stripes}: ‘{How} {Possibly}’ vs. ‘{How} {Actually}’ {Model} {Explanations}},
	volume = {97},
	issn = {0026-9662},
	shorttitle = {How the {Tiger} {Bush} {Got} its {Stripes}},
	url = {https://doi.org/10.5840/monist201497321},
	doi = {10.5840/monist201497321},
	number = {3},
	urldate = {2026-02-20},
	journal = {The Monist},
	author = {Bokulich, Alisa},
	month = jul,
	year = {2014},
	pages = {321--338},
}

@book{BrandonBrandon+2014,
	address = {Princeton},
	title = {Adaptation and environment},
	isbn = {978-1-4008-6066-1},
	url = {https://doi.org/10.1515/9781400860661},
	doi = {doi:10.1515/9781400860661},
	urldate = {2026-03-19},
	publisher = {Princeton University Press},
	author = {Brandon, Robert N. and Brandon, Robert N.},
	year = {2014},
}

@book{eganDeflatingMentalRepresentation2025a,
	title = {Deflating {Mental} {Representation} ({The} {Jean} {Nicod} {Lectures})},
	publisher = {MIT Press (open access)},
	author = {Egan, Frances},
	year = {2025},
}

@misc{chenPersonaVectorsMonitoring2025,
	title = {Persona {Vectors}: {Monitoring} and {Controlling} {Character} {Traits} in {Language} {Models}},
	shorttitle = {Persona {Vectors}},
	url = {http://arxiv.org/abs/2507.21509},
	doi = {10.48550/arXiv.2507.21509},
	urldate = {2026-02-03},
	publisher = {arXiv},
	author = {Chen, Runjin and Arditi, Andy and Sleight, Henry and Evans, Owain and Lindsey, Jack},
	month = sep,
	year = {2025},
	note = {arXiv:2507.21509 [cs]},
}

@article{shmueliExplainPredict2010,
	title = {To {Explain} or to {Predict}?},
	volume = {25},
	issn = {0883-4237},
	url = {https://projecteuclid.org/journals/statistical-science/volume-25/issue-3/To-Explain-or-to-Predict/10.1214/10-STS330.full},
	doi = {10.1214/10-STS330},
	language = {en},
	number = {3},
	urldate = {2026-03-18},
	journal = {Statistical Science},
	author = {Shmueli, Galit},
	month = aug,
	year = {2010},
}

@inproceedings{weidingerTaxonomyRisksPosed2022,
	address = {Seoul Republic of Korea},
	title = {Taxonomy of {Risks} posed by {Language} {Models}},
	isbn = {978-1-4503-9352-2},
	url = {https://dl.acm.org/doi/10.1145/3531146.3533088},
	doi = {10.1145/3531146.3533088},
	language = {en},
	urldate = {2026-03-21},
	booktitle = {2022 {ACM} {Conference} on {Fairness} {Accountability} and {Transparency}},
	publisher = {ACM},
	author = {Weidinger, Laura and Uesato, Jonathan and Rauh, Maribeth and Griffin, Conor and Huang, Po-Sen and Mellor, John and Glaese, Amelia and Cheng, Myra and Balle, Borja and Kasirzadeh, Atoosa and Biles, Courtney and Brown, Sasha and Kenton, Zac and Hawkins, Will and Stepleton, Tom and Birhane, Abeba and Hendricks, Lisa Anne and Rimell, Laura and Isaac, William and Haas, Julia and Legassick, Sean and Irving, Geoffrey and Gabriel, Iason},
	month = jun,
	year = {2022},
	pages = {214--229},
}

@misc{lazaridouMindGapAssessing2021,
	title = {Mind the {Gap}: {Assessing} {Temporal} {Generalization} in {Neural} {Language} {Models}},
	shorttitle = {Mind the {Gap}},
	url = {http://arxiv.org/abs/2102.01951},
	doi = {10.48550/arXiv.2102.01951},
	urldate = {2026-03-21},
	publisher = {arXiv},
	author = {Lazaridou, Angeliki and Kuncoro, Adhiguna and Gribovskaya, Elena and Agrawal, Devang and Liska, Adam and Terzi, Tayfun and Gimenez, Mai and d'Autume, Cyprien de Masson and Kocisky, Tomas and Ruder, Sebastian and Yogatama, Dani and Cao, Kris and Young, Susannah and Blunsom, Phil},
	month = oct,
	year = {2021},
	note = {arXiv:2102.01951 [cs]},
}

@misc{weidingerEthicalSocialRisks2021,
	title = {Ethical and social risks of harm from {Language} {Models}},
	url = {http://arxiv.org/abs/2112.04359},
	doi = {10.48550/arXiv.2112.04359},
	urldate = {2026-03-21},
	publisher = {arXiv},
	author = {Weidinger, Laura and Mellor, John and Rauh, Maribeth and Griffin, Conor and Uesato, Jonathan and Huang, Po-Sen and Cheng, Myra and Glaese, Mia and Balle, Borja and Kasirzadeh, Atoosa and Kenton, Zac and Brown, Sasha and Hawkins, Will and Stepleton, Tom and Biles, Courtney and Birhane, Abeba and Haas, Julia and Rimell, Laura and Hendricks, Lisa Anne and Isaac, William and Legassick, Sean and Irving, Geoffrey and Gabriel, Iason},
	month = dec,
	year = {2021},
	note = {arXiv:2112.04359 [cs]},
}

@misc{guestCriticalArtificialIntelligence2025,
	title = {Critical {Artificial} {Intelligence} {Literacy} for {Psychologists}},
	url = {https://osf.io/preprints/psyarxiv/dkrgj_v1/},
	doi = {10.31234/osf.io/dkrgj_v1},
	urldate = {2026-03-21},
	publisher = {PsyArXiv},
	author = {Guest, Olivia and van Rooij, Iris},
	month = oct,
	year = {2025},
}

@misc{groeneveldOLMoAcceleratingScience2024,
	title = {{OLMo}: {Accelerating} the {Science} of {Language} {Models}},
	shorttitle = {{OLMo}},
	url = {http://arxiv.org/abs/2402.00838},
	doi = {10.48550/arXiv.2402.00838},
	urldate = {2025-05-30},
	publisher = {arXiv},
	author = {Groeneveld, Dirk and Beltagy, Iz and Walsh, Pete and Bhagia, Akshita and Kinney, Rodney and Tafjord, Oyvind and Jha, Ananya Harsh and Ivison, Hamish and Magnusson, Ian and Wang, Yizhong and Arora, Shane and Atkinson, David and Authur, Russell and Chandu, Khyathi Raghavi and Cohan, Arman and Dumas, Jennifer and Elazar, Yanai and Gu, Yuling and Hessel, Jack and Khot, Tushar and Merrill, William and Morrison, Jacob and Muennighoff, Niklas and Naik, Aakanksha and Nam, Crystal and Peters, Matthew E. and Pyatkin, Valentina and Ravichander, Abhilasha and Schwenk, Dustin and Shah, Saurabh and Smith, Will and Strubell, Emma and Subramani, Nishant and Wortsman, Mitchell and Dasigi, Pradeep and Lambert, Nathan and Richardson, Kyle and Zettlemoyer, Luke and Dodge, Jesse and Lo, Kyle and Soldaini, Luca and Smith, Noah A. and Hajishirzi, Hannaneh},
	month = jun,
	year = {2024},
	note = {arXiv:2402.00838 [cs]},
}

@misc{jackson_2024_13914560,
	title = {Public {AI}: {Infrastructure} for the common good},
	url = {https://doi.org/10.5281/zenodo.13914560},
	doi = {10.5281/zenodo.13914560},
	publisher = {Public AI Network},
	author = {Jackson, Brandon and Cavello, B and Devine, Flynn and Garcia, Nick and Klein, Samuel J. and Krasodomski, Alex and Tan, Joshua and Tursman, Eleanor},
	month = aug,
	year = {2024},
}

@book{morganModelsMediatorsPerspectives1999,
	address = {Cambridge},
	series = {Ideas in {Context}},
	title = {Models as {Mediators}: {Perspectives} on {Natural} and {Social} {Science}},
	isbn = {978-0-521-65097-7},
	shorttitle = {Models as {Mediators}},
	url = {https://www.cambridge.org/core/books/models-as-mediators/FBB3EA4AECAF824AD6F1E6C650CAE3AE},
	doi = {10.1017/CBO9780511660108},
	urldate = {2026-03-19},
	publisher = {Cambridge University Press},
	editor = {Morgan, Mary S. and Morrison, Margaret},
	year = {1999},
}

@book{duhemAimStructurePhysical1954,
	title = {The aim and structure of physical theory},
	volume = {1},
	publisher = {Princeton University Press},
	author = {Duhem, Pierre Maurice Marie},
	year = {1954},
	note = {Pages: 85-87},
}

@book{quineLogicalPointView1953,
	address = {Cambridge},
	title = {From a {Logical} {Point} of {View}},
	publisher = {Harvard University Press},
	author = {Quine, Willard Van Orman},
	year = {1953},
}

\newpage
\appendix

\section{Examples}
\label{sec:ChecklistExamples}
In this section we step through Figures \ref{fig:ConwayChecklist}-\ref{fig:AnasaziChecklist}, which contain example checklists filled out for each level.

\begin{figure}[ht!] 
        \centering
        \vspace*{\fill}
    \begin{plausibilitybox}{Conway's Game of Life
        }
    \small 
    This scale grades the model’s potential as a plausible explanation for the target phenomenon. While not every model’s goal is to be explanatory, it is important to clarify when it is appropriate.
    \chksection{Level 0: Simulation ($S$) -- \textit{Sandbox/Toy Model}}
    \begin{itemize}[leftmargin=1.5em, label=$\square$]
        \item[\cbox] \textbf{Simulation.} Is the simulation (S) defined, including environments, agents, and update rules?
    \end{itemize}
    \noindent {\color{red} The rules of the game are detailed: At each timestep, \textit{``1. Every counter with two or three neighboring counters survives for the next generation. 2. Each counter with four or more neighbors dies (is removed) from overpopulation. Every counter with one neighbor or none dies from isolation. Each empty cell adjacent to exactly three neighbors--no more, no fewer--is a birth cell. A counter is placed on it at the next move.''}}
    
    \chksection{Level 1: Target Phenomenon ($T$) -- \textit{Phenomenal Model}}
    \begin{itemize}[leftmargin=1.5em, label=$\square$]
        \item \textbf{Defined Target Phenomenon.} Is a target phenomenon ($T$) operationalized (e.g. as statistical patterns, human annotations/observations, etc.)?
        \item \textbf{Generative Sufficiency.} Can the simulation ($S$) successfully generate the patterns described in ($T$)?
        \item[\cbox] \textbf{Reproducibility.} Does the reproducibility of the simulation (seeds, API versions, consideration of proprietary prompt injections or version changes, inherent stochasticity, etc.) match the reproducibility goals of the modeler or the field they are working in (sensitivity analysis requirements, etc.)?
    \end{itemize}
    {\color{red} While Conway selected his rules to try and make the behavior of populations in his sim unpredictable, for the reader there is no operationalized pattern that the simulation is intended to reproduce, his simulation described as a ``solitaire'' \cite{gardnerMathematicalGames1970}.}
    
    \chksection{Level 2: Intent \& Mapping ($I$) -- \textit{How-Possibly Model}}
    \begin{itemize}[leftmargin=1.5em, label=$\square$]
        \item \textbf{Simulation Contribution.} Is the simulation's use case understood (e.g., predictive, exploratory, illustrative, explanatory, etc.)?
        \item \textbf{Falsifiable Hypotheses.} Does there exist a hypothesis for how the target phenomenon $T$ arises from components in the model?
        \item \textbf{Mechanism Mapping.} Is there an Intent $I$ (implicit or explicit mapping) which connects components of the simulation ($S$) to the hypothesized `real-world' mechanisms of ($T$)?
    \end{itemize}
    {\color{red} Due to the lack of $T$, the simulation does not describe any mechanisms. Mechanisms are defined relative to a phenomenon (see `Glennan's Law' \cite{glennanMechanismsNatureCausation1996, craverMechanismsScience2024}).}
    
    \chksection{Level 3: Evidence ($E$) -- \textit{Plausible Model (if validated)}}
    \begin{itemize}[leftmargin=1.5em, label=$\square$]
        \item \textbf{Evidence Exists.} Is the model supported by some evidence $E$?
        \item \textbf{Relevance.} Is $E$ directed towards the claims made in $I$? Could $E$, in principle, disconfirm the hypotheses in $I$?
    \end{itemize}
    {\color{red}Since $E$ requires $T$ and $I$, the model does not pass the Level 3 requirement.}
    
    \vspace{4pt}
    \chksection{\textit{\textbf{Conclusion:} Based on the checklist, the model is classified as a \textbf{Level 0 Toy} model.}}
    
    \end{plausibilitybox}
    \caption{Example for Level 0: The Mechanism Plausibility Scale applied to an implementation of Conway's Game of Life \cite{gardnerMathematicalGames1970}.}
        \label{fig:ConwayChecklist}
\end{figure}

\clearpage

\begin{figure}[ht!] 
        \centering
        \vspace*{\fill}
    \begin{plausibilitybox}{Game Theory LLM Benchmarking Sim}
    \small 
    This scale grades the model’s potential as a plausible explanation for the target phenomenon. While not every model’s goal is to be explanatory, it is important to clarify when it is appropriate.
    \chksection{Level 0: Simulation ($S$) -- \textit{Sandbox/Toy Model}}
    \begin{itemize}[leftmargin=1.5em, label=$\square$]
        \item[\cbox] \textbf{Simulation.} Is the simulation (S) defined, including environments, agents, and update rules?
    \end{itemize}
    \noindent {\color{red} $S$ includes the codebase for the simulation, which is an implementation of games such as Prisoner's Dilemma, Texas Hold'em, Staghunt, etc. played by different LLM Agents against each other.}
    
    \chksection{Level 1: Target Phenomenon ($T$) -- \textit{Phenomenal Model}}
    \begin{itemize}[leftmargin=1.5em, label=$\square$]
        \item[\cbox] \textbf{Defined Target Phenomenon.} Is a target phenomenon ($T$) operationalized (e.g. as statistical patterns, human annotations/observations, etc.)?
        \item[\cbox] \textbf{Generative Sufficiency.} Can the simulation ($S$) successfully generate the patterns described in ($T$)?
        \item[\cbox] \textbf{Reproducibility.} Does the reproducibility of the simulation (e.g. seeds, API versions, consideration of proprietary prompt injections or version changes, inherent stochasticity) match the reproducibility goals of the modeler or the field they are working in (sensitivity analysis requirements, etc.)?
    \end{itemize}
    {\color{red} We aim to benchmark the behavior of different LLMs to see if they will solve game-theoretic scenarios, and if they produce optimal or well-known game-theoretic behavior (e.g., Tit-for-Tat, Nash Equilibrium); their execution of these strategies we will consider $T$. After running the simulation, we find that models with a larger parameter space tend to exhibit more aggressive behavior.}
    
    \chksection{Level 2: Intent \& Mapping ($I$) -- \textit{How-Possibly Model}}
    \begin{itemize}[leftmargin=1.5em, label=$\square$]
        \item \textbf{Simulation Contribution.} Is the simulation's use case understood (e.g., predictive, exploratory, illustrative, explanatory, etc.)?
        \item \textbf{Falsifiable Hypotheses.} Does there exist a hypothesis for how the target phenomenon $T$ arises from components in the model?
        \item \textbf{Mechanism Mapping.} Is there an Intent $I$ (implicit or explicit mapping) which connects components of the simulation ($S$) to the hypothesized `real-world' mechanisms of ($T$)?
    \end{itemize}
    {\color{red} We do not suggest the mechanisms behind why some models play different styles of games than others. However, our findings do find a correlation between the aggressiveness of players and their parameter count.}
    
    \chksection{Level 3: Evidence ($E$) -- \textit{Plausible Model (if validated)}}
    \begin{itemize}[leftmargin=1.5em, label=$\square$]
        \item \textbf{Evidence Exists.} Is the model supported by some evidence $E$?
        \item \textbf{Relevance.} Is $E$ directed towards the claims made in $I$? Could $E$, in principle, disconfirm the hypotheses in $I$?
    \end{itemize}
    {\color{red} Since $E$ requires $I$, the model does not pass the Level 3 requirement.}
    
    \vspace{4pt}
    \chksection{\textit{\textbf{Conclusion:} Based on the checklist, the model is classified as a \textbf{Level 1 Phenomenal} model.}}
    
    \end{plausibilitybox}
    \caption{Example for Level 1: The Mechanism Plausibility Scale applied to a fabricated game theory paper.}
        \label{fig:GameTheoryChecklist}
\end{figure}

\clearpage

\begin{figure*}[ht!] 
    \centering
    \vspace*{\fill}
    \begin{plausibilitybox}{Example Schelling's Model of Segregation}
    \small 
    This scale grades the model’s potential as a plausible explanation for the target phenomenon. While not every model’s goal is to be explanatory, it is important to clarify when it is appropriate.
    \chksection{Level 0: Simulation ($S$) -- \textit{Sandbox/Toy Model}}
    \begin{itemize}[leftmargin=1.5em, label=$\square$]
        \item[\cbox] \textbf{Simulation.} Is the simulation (S) defined, including environments, agents, and update rules?
    \end{itemize}
    \noindent {\color{red} $S$ is defined as a grid of agents who move to adjacent empty spots if the percentage of their own color neighbors falls below a certain threshold (formalized in full paper).}

    \chksection{Level 1: Target Phenomenon ($T$) -- \textit{Phenomenal Model}}
    \begin{itemize}[leftmargin=1.5em, label=$\square$]
        \item[\cbox] \textbf{Defined Target Phenomenon.} Is a target phenomenon ($T$) operationalized (e.g. as statistical patterns, human annotations/observations, etc.)?
        \item[\cbox] \textbf{Generative Sufficiency.} Can the simulation ($S$) successfully generate the patterns described in ($T$)?
        \item[\cbox] \textbf{Reproducibility.} Is the simulation reproducible?
    \end{itemize}
    {\color{red} The target $T$ is the emergence of macro-level clustering, which represents residential segregation. The simulation is generatively sufficient as it always produces segregated neighborhoods.}

    \chksection{Level 2: Intent \& Mapping ($I$) -- \textit{How-Possibly Model}}
    \begin{itemize}[leftmargin=1.5em, label=$\square$]
        \item[\cbox] \textbf{Simulation Contribution.} Is the simulation's use case understood (e.g., predictive, exploratory, illustrative, explanatory, etc.)?
        \item[\cbox] \textbf{Falsifiable Hypotheses.} Does there exist a hypothesis for how the target phenomenon $T$ arises from components in the model?
        \item[\cbox] \textbf{Mechanism Mapping.} Is there an Intent $I$ (implicit or explicit mapping) which connects components of the simulation ($S$) to the hypothesized `real-world' mechanisms of ($T$)?
    \end{itemize}
    {\color{red} Our Intent $I$ is to demonstrate the feasibility of the hypothesis that extreme individual prejudice is not necessary for macro-segregation to occur. Mild preferences for similar neighbors are a sufficient mechanism. The agents' movement rules map to a simplification of the real-world mechanism of residents relocating based on neighborhood composition.}

    \chksection{Level 3: Evidence ($E$) -- \textit{Plausible Model (if validated)}}
    \begin{itemize}[leftmargin=1.5em, label=$\square$]
        \item \textbf{Evidence Exists.} Is the model supported by some evidence $E$?
        \item \textbf{Relevance.} Is $E$ directed towards the claims made in $I$? Could $E$, in principle, disconfirm the hypotheses in $I$?
    \end{itemize}
    {\color{red} The threshold parameters in our model are abstract and not derived from specific empirical data (e.g. census surveys). It proposes a ``how-possibly'' mechanism.}

    \vspace{4pt}
    \chksection{\textit{\textbf{Conclusion:} Based on the checklist, the model is classified as a \textbf{Level 2 How-Possibly} model.}}

    \end{plausibilitybox}
    \caption{Example for Level 2: The Mechanism Plausibility Scale applied to Schelling's Model of Segregation \cite{schellingModelsSegregation1969}}
    \label{fig:SchellingChecklist}
    \vspace*{\fill}
\end{figure*}

\clearpage

\begin{figure*}[ht!] 
    \centering
    \vspace*{\fill}
    \begin{plausibilitybox}{Example Anasazi Model (Epstein et al.)}
    \small 
    This scale grades the model’s potential as a plausible explanation for the target phenomenon. While not every model’s goal is to be explanatory, it is important to clarify when it is appropriate.
    \chksection{Level 0: Simulation ($S$) -- \textit{Sandbox/Toy Model}}
    \begin{itemize}[leftmargin=1.5em, label=$\square$]
        \item[\cbox] \textbf{Simulation.} Is the simulation (S) defined, including environments, agents, and update rules?
    \end{itemize}
    \noindent {\color{red} $S$ is an agent-based model that simulates the Long House Valley environment using rules for annual agricultural productivity, and defines agent behaviors based on specific rules for nutritional needs, household size, and reproduction rates. The details of these parameters, as well as the simulation code, can be found in our GitHub repository.}

    \chksection{Level 1: Target Phenomenon ($T$) -- \textit{Phenomenal Model}}
    \begin{itemize}[leftmargin=1.5em, label=$\square$] 
        \item[\cbox] \textbf{Defined Target Phenomenon.} Is a target phenomenon ($T$) operationalized (e.g. as statistical patterns, human annotations/observations, etc.)?
        \item[\cbox] \textbf{Generative Sufficiency.} Can the simulation ($S$) successfully generate the patterns described in ($T$)?
        \item[\cbox] \textbf{Reproducibility.} Is the simulation reproducible?
    \end{itemize}
    {\color{red} The target $T$ is the historical population dynamics and settlement patterns of the Long House Valley from 800 AD to 1350 AD. The simulation reproduces the population crash and abandonment of the valley.}

    \chksection{Level 2: Intent \& Mapping ($I$) -- \textit{How-Possibly Model}}
    \begin{itemize}[leftmargin=1.5em, label=$\square$]
        \item[\cbox] \textbf{Simulation Contribution.} Is the simulation's use case understood (e.g., predictive, exploratory, illustrative, explanatory, etc.)?
        \item[\cbox] \textbf{Falsifiable Hypotheses.} Does there exist a hypothesis for how $T$ arises?
        \item[\cbox] \textbf{Mechanism Mapping.} Is there an Intent $I$ (implicit or explicit mapping) which connects components of the simulation ($S$) to the hypothesized `real-world' mechanisms of ($T$)?
    \end{itemize}
    {\color{red} The intent ($I$) maps the simulation's rules (agricultural yield vs. caloric needs) to the real-world mechanisms of the environmental factors affecting population growth. Therefore it hypothesizes that environmental shifts were the primary driver.}

    \chksection{Level 3: Evidence ($E$) -- \textit{Plausible Model (if validated)}}
    \begin{itemize}[leftmargin=1.5em, label=$\square$]
        \item[\cbox] \textbf{Evidence Exists.} Is the model supported by some evidence $E$?
        \item[\cbox] \textbf{Relevance.} Is $E$ directed towards the claims made in $I$? Could $E$, in principle, disconfirm the hypotheses in $I$?
    \end{itemize}
    {\color{red} The model is constrained by empirical Evidence $E$, which includes agricultural productivity reconstructed using paleoenvironmental data from tree rings, soil analysis, and geology. Each agent's nutritional needs and household sizes are derived from anthropological studies of Puebloan peoples.}

    \vspace{4pt}
    \chksection{\textit{\textbf{Conclusion:} Based on the checklist, the model is classified as a \textbf{Level 3 Plausible} model.}}

    \end{plausibilitybox}
    \caption{Example for Level 3: The Mechanism Plausibility Scale applied to the Artificial Anasazi Model \cite{epsteinGenerativeSocialScience2006}}
    \label{fig:AnasaziChecklist}
    \vspace*{\fill}
\end{figure*}

\clearpage

\section{Calibrating the Mechanism Plausibility Scale}
\label{sec:CalibrationAppendix}
This section goes through the process of how we iterated on the scale.

\subsection{Calibration}
The Mechanism Plausibility Scale was refined through double-blinded review processes involving papers drawn from a systematic review of LLM-based social simulations by Larooij et al. \cite{larooijLargeLanguageModels2025}. Each paper was independently evaluated by two reviewers who assigned scores before entering a reconciliation phase. After going through two rounds of calibration, multiple ambiguities still remained about how the scale should be applied. This led to further rework where we decided it would be more appropriate to reframe the scale as a practical checklist format.

\subsection{Round 1 Calibration}
\label{subsec:AppliedReviewDesign}
\subsubsection{Paper Selection and Review Process}

We tested early versions of the scale on the first 15 out of 35 papers from Larooij et al.'s systematic review \cite{larooijLargeLanguageModels2025}.
We chose to evaluate papers from  Larooij et al.'s review because we found that their inclusion criteria was heavily aligned with our own research interests.
In particular, the requirements that the ABM uses an LLM as the basis for their agents, there are multiple interacting agents, and that the LLMs were seen to be simulating human behavior, were all aligned with our own conceptions of LLM-ABM social simulation. A copy of the exact queries they used can be found in our Appendix at \ref{quote:ABMQuery}.

Each paper was reviewed and evaluated by two reviewers; their task comprised of two phases: (1) the evaluation period, and (2) the reconciliation period.
During the evaluation period, each reviewer would read a paper and assign it a score before moving onto the next paper. 
The reviewers were blinded to the scores and sentiments of the other reviewer until the reconciliation period began.

\subsubsection{Inter-Rater Reliability}

To assess the reliability of the Mechanism Plausibility Scale, we calculated the inter-rater reliability using a weighted kappa ($k_w$) with quadratic weights, suitable for our ordinal scale. As mentioned before, the first structured round of applying the scale to a focused body of literature revealed that it was challenging for even two researchers to apply the scale consistently; we found ambiguities in the rating guidelines and confusion with the nested nature of the simulations through the reconciliation process.

\newpage
\subsubsection{Round 1 results of the applied review}
\begin{table}[h!]
\centering
\begin{tabular}{|l|c|c|}
\hline
\multicolumn{3}{|c|}{\textbf{Round 1}} \\
\hline
~ & \multicolumn{2}{c|}{\textbf{Reviewer Scores}} \\
\hline
\textbf{Shortened Title} & \textbf{A} & \textbf{B} \\
\hline
Generative Agents \cite{parkGenerativeAgentsInteractive2023}
    & 3 & 1 \\
\hline
WarAgent \cite{huaWarPeaceWarAgent2024}
    & 3 & 1 \\
\hline
Social Simulacra \cite{parkSocialSimulacraCreating2022}
    & 3 & 1 \\
\hline
S3 \cite{gaoS3SocialnetworkSimulation2025}
    & 0 & 1 \\
\hline
Scale-Free Networks \cite{marzoEmergenceScaleFreeNetworks2023}
    & 3 & 2 \\
\hline
Humanoid Agents \cite{wangHumanoidAgentsPlatform2023}
    & 3 & 2 \\
\hline
LyfeAgents \cite{kaiyaLyfeAgentsGenerative2023}
    & 3 & 0  \\
\hline
Collaboration \cite{zhangExploringCollaborationMechanisms2024}
    & 3 & 2 \\
\hline
AgentVerse \cite{chenAgentVerseFacilitatingMultiAgent2023}
    & 3 & 0 \\
\hline
Epidemic Modeling \cite{williamsEpidemicModelingGenerative2023}
    & 3 & 2 \\
\hline
Project Sid \cite{alProjectSidManyagent2024}
    & 3 & 1 \\
\hline
Theory of Mind \cite{liTheoryMindMultiAgent2023}
    & 3 & 1 \\
\hline
News Diffusion \cite{liLargeLanguageModeldriven2024}
    & 0 & 1 \\
\hline
SpeechAgents \cite{zhangSpeechAgentsHumanCommunicationSimulation2024}
    & 1 & 1 \\
\hline
Fake News Propagation \cite{liuSkepticismAcceptanceSimulating2024}
    & 1 & 1 \\
\hline
\end{tabular}
\caption{Levels of the Mechanism Plausibility Scale assigned to each paper by Reviewers A and B during the blinded first round of evaluation, along with its assigned level after unblinding and reconciliation.}
\label{tab:AppliedReview1}
\end{table}

The results of the first round are shown in Table \ref{tab:AppliedReview1}.
Through the reconciliation process, we found that our scale's rating guidelines and definitions were too ambiguous to handle the operationalization gaps discussed in Section \ref{subsec:Reflections}.

Notably, we also found that many papers conflated Agent-level functionality with ABM-level plausibility.
A paper might have provided high-quality experimental evidence for its Agent-level social simulation (the lowest level in Figure \ref{fig:CraverDiagram}), but then they implicitly treated that as sufficient evidence for the claims made about the emergent social phenomenon ($T$) observed at the ABM-level, which is a category error.
Work that shows the mechanism plausibility of Agent-level phenomena does not necessarily translate to the mechanism plausibility of ABM-level phenomena. This remains to be shown and must be argued for in the modeler's Intent, with further Evidence provided at the ABM-level.

Papers had this problem to varying degrees, but the ones that we flagged particularly were:
\begin{itemize}
\item S3 \cite{gaoS3SocialnetworkSimulation2025}, which conflated (the LLM agent's capacity to simulate the social media posts of an individual, with matching estimated emotions and attitudes) with (their ABM's ability to simulate realistic social media phenomena, like opinion dynamics or information cascades).
\item News Diffusion \cite{liLargeLanguageModeldriven2024}, which conflated their (LLM agent's ability to share news based on personality traits and friend connections), with their (ABM's capacity to produce realistic fake news diffusion patterns).
\item SpeechAgents \cite{zhangSpeechAgentsHumanCommunicationSimulation2024}, which conflated (their LLM agent's ability to generate text-to-speech outputs, which was successfully transcribed back into similar text) with (their ABM's ability to simulate realistic, emergent human communication dynamics and social interaction patterns at the group level).
\item Fake News Propagation \cite{liuSkepticismAcceptanceSimulating2024}, which conflated (an LLM agent's capability to reason about, reflect on, and share fake news) with (their ABM's ability to mechanistically explain realistic fake news propagation dynamics and the emergence of collective opinion patterns).
\end{itemize}

The presence of these nested simulations was particularly difficult to evaluate with our scale, as it was difficult to identify the target phenomenon $T$, and also difficult to identify its operationalization. The ambiguous evaluations between Agent and ABM phenomena led to a split between the reviewer's perceptions and resulted in an initial quadratic weighted kappa score of 0.207.

\subsection{Round 2 Calibration}
\label{subsubsec:AppliedReviewResults}
After uncovering the gaps in operationalization (also discussed in Section \ref{subsec:Reflections}), in the second round the reviewers were to assign two scores: one for the ABM-level target phenomena, and one for the Agent-level target phenomena. In addition to the clarified scoring guidelines, the same reviewers were also used in both rounds. and so the increase in reliability may be partially attributable to the training received as part of the first round. 
 
As shown in Table \ref{tab:KappaReview}, the quadratic weighted kappa for the ABM ratings rose to 0.255, and the score for the Agent ratings reached 0.503. We posit that the score difference between Agent and ABM also comes from papers from the literature review operationalizing the Agent-level phenomena (implicitly) more in-depth compared to the ABM-level. These results are shown in Table \ref{tab:AppliedReview2}.

\begin{table}[h]
\centering
\begin{tabular}{|l|c|c|c|c|c|c|}
\hline
\multicolumn{7}{|c|}{\textbf{Round 2}} \\
\hline
~ & \multicolumn{6}{c|}{\textbf{Reviewer Scores}} \\
\hline
~ & \multicolumn{3}{|c|}{\textbf{ABM}} & \multicolumn{3}{|c|}{\textbf{Agent}} \\
\hline
\textbf{Shortened Title}
    & \textbf{A} & \textbf{B} & \textbf{R}
    & \textbf{A} & \textbf{B} & \textbf{R}
    \\
\hline
Generative Agents \cite{parkGenerativeAgentsInteractive2023}
    & 3 & 1 & 3
    & 3 & 3 & 3
    \\
\hline
WarAgent \cite{huaWarPeaceWarAgent2024}
    & 3 & 1 & 3
    & 3 & 1 & 3
    \\
\hline
Social Simulacra \cite{parkSocialSimulacraCreating2022}
    & 1 & 1 & 1
    & 3 & 3 & 3
    \\
\hline
S3 \cite{gaoS3SocialnetworkSimulation2025}
    & 0 & 0 & 0
    & 1 & 2 & 1
    \\
\hline
Scale-Free Networks \cite{marzoEmergenceScaleFreeNetworks2023}
    & 3 & 3 & 3
    & 1 & 1 & 1
    \\
\hline
Humanoid Agents \cite{wangHumanoidAgentsPlatform2023}
    & 1 & 0 & 1
    & 3 & 3 & 3
    \\
\hline
LyfeAgents \cite{kaiyaLyfeAgentsGenerative2023}
    & 1 & 1 & 1
    & 3 & 1 & 3
    \\
\hline
Collaboration \cite{zhangExploringCollaborationMechanisms2024}
    & 3 & 2 & 3
    & 3 & 3 & 3
    \\
\hline
AgentVerse \cite{chenAgentVerseFacilitatingMultiAgent2023}
    & 3 & 2 & 3
    & 3 & 1 & 3
    \\
\hline
Epidemic Modeling \cite{williamsEpidemicModelingGenerative2023}
    & 3 & 2 & 3
    & 3 & 2 & 3
    \\
\hline
Project Sid \cite{alProjectSidManyagent2024}
    & 3 & 2 & 3
    & 3 & 1 & 3
    \\
\hline
Theory of Mind \cite{liTheoryMindMultiAgent2023}
    & 1 & 2 & 2
    & 3 & 2 & 3
    \\
\hline
News Diffusion \cite{liLargeLanguageModeldriven2024}
    & 3 & 2 & 3
    & 2 & 1 & 2
    \\
\hline
SpeechAgents \cite{zhangSpeechAgentsHumanCommunicationSimulation2024}
    & 1 & 0 & 1
    & 3 & 3 & 3
    \\
\hline
Fake News Propagation \cite{liuSkepticismAcceptanceSimulating2024}
    & 3 & 2 & 3
    & 3 & 2 & 3
    \\
\hline
\end{tabular}
\caption{Levels of the Mechanism Plausibility Scale assigned to each paper by Reviewers A and B in the blinded second round of the literature review, along with its assigned level after unblinding and reconciliation.}
\label{tab:AppliedReview2}
\end{table}


\begin{table}[h]
\centering
\begin{tabular*}{\columnwidth}{@{\extracolsep{\fill}}lcc}
\toprule
\textbf{Dataset} & \textbf{Weighted Kappa ($k_w$)} & \textbf{Interpretation} \\
\midrule
Round 1 & 0.207 & Fair Agreement \\
Round 2 (Agent) & 0.503 & Moderate Agreement \\
Round 2 (ABM) & 0.255 & Fair Agreement \\
\bottomrule
\end{tabular*}
\caption{Summary of weighted kappa ($k_w$) scores using quadratic weights. The ordinal measure was used because the plausibility scale has ordered categories. The interpretation comes from Landis and Koch \cite{landisMeasurementObserverAgreement1977}.}
\label{tab:KappaReview}
\end{table}

In the second round of evaluations, we found that our definition of the Evidence (as stated in Section \ref{subsec:Level3}) was too broad, allowing for many papers to reach a Level 3 on our scale, regardless of the quality of their Evidence.
Difficulties in ignoring a paper's perceived evidence quality led to large differences between reviewer scores at the ABM-level.

\subsection{Queries for LLM ABM-related papers}
The 15 papers from our applied study are from Larooij et al. \cite{larooijLargeLanguageModels2025}:
\begin{quote}
    \texttt{TITLE-ABS-KEY ( ( "generative social simulation" ) OR ( "generative
agent-based model*" ) OR ( "agent-based simulation" AND "generative
AI" ) OR ( "LLM*" AND "agent-based model*" ) OR ( "large language
model*" AND "ABM" ) OR ( "foundation model*" AND "ABM" ) OR ( "multiagent system*" AND "generative AI" ) OR ( "generative agent*" )
OR ( "social simulation" AND "LLM*" ) OR ( "large language model-based
agents" ) )}
\label{quote:ABMQuery}
\end{quote}

We also used Asta\footnote{\url{https://asta.allen.ai/chat}}, an AI research paper search engine built on Semantic Scholar, for paper discovery. Other LLM-ABM papers were from previous knowledge of the authors or related work.

\section{Additional Philosophy of Science Background}
\subsection{Clarification of Model Targets $T$}
According to a description by Weisberg, $T$ does not have to be a particular or `real' phenomenon, and some models might not even have a $T$; the target could be \textit{particular}, \textit{generic}, or even \textit{hypothetical} \cite{weisbergSimulationSimilarityUsing2013}. As an example, Graebner gives an account of how the target of Schelling's segregation model \cite{schellingModelsSegregation1969} is a \textit{generalized} target, where it represents an abstract city and its arguments can be applied generically rather than to a particular city \cite{graebnerHowRelateModels2018, weisbergSimulationSimilarityUsing2013}. $T$ can also be \textit{hypothetical}; for example, Fisher's three-sex population simulation \cite{fisherGeneticalTheoryNatural1999} shows how a population with three sexes comes with large costs compared to those with only two and is a possible explanation for why three-sex populations do not exist in reality \cite{graebnerHowRelateModels2018, weisbergWhoModeler2007}.

\end{document}